\documentclass[%
 aip,jcp,
 amsmath,amssymb,
 reprint,%
]{revtex4-1}

\usepackage{graphicx}
\usepackage{dcolumn}
\usepackage{bm}
\usepackage{multirow}
\usepackage[utf8]{inputenc}
\usepackage[T1]{fontenc}
\usepackage{mathptmx}

\begin{document}

\preprint{AIP/123-QED}

\title{Computational method for highly-constrained molecular dynamics of rigid bodies: coarse-grained simulation of auxetic two-dimensional protein crystals}

\author{Jorge A. Campos-Gonzalez-Angulo}
 \altaffiliation{Department of Chemistry and Biochemistry, University of California San Diego, La Jolla, California 92093, United States}
\author{Garret Wiesehan}%
\affiliation{Department of Chemistry and Biochemistry, University of California San Diego, La Jolla, California 92093, United States}%
\author{Raphael F. Ribeiro}
\affiliation{Department of Chemistry and Biochemistry, University of California San Diego, La Jolla, California 92093, United States}%
\author{Joel Yuen-Zhou}%
 \email{joelyuen@ucsd.edu}
 \homepage{http://yuenzhougroup.ucsd.edu/}
\affiliation{Department of Chemistry and Biochemistry, University of California San Diego, La Jolla, California 92093, United States}%

\begin{abstract}
The increasing number of protein-based metamaterials demands reliable and efficient theoretical and computational methods to study the physicochemical properties they may display. In this regard, we develop a simulation strategy based on Molecular Dynamics (MD) that addresses the geometric degrees of freedom of an auxetic two-dimensional protein crystal. This model consists of a network of impenetrable rigid squares linked through massless rigid rods. Our MD methodology extends the well-known protocols SHAKE and RATTLE to include highly non-linear holonomic and non-holonomic constraints, with emphasis on collision detection and response between anisotropic rigid bodies. The presented method enables the simulation of long-time dynamics with reasonably large time-steps.  The data extracted from the simulations allow the characterization of the dynamical correlations featured by the protein subunits, which show a persistent motional interdependence across the array. On the other hand, non-holonomic constraints (collisions between subunits) increase the number of inhomogeneous deformations of the network, thus driving it away from an isotropic response. Our work provides the first long-timescale simulation of the dynamics of protein crystals and offers insights into promising mechanical properties afforded by these materials.
\end{abstract}

\maketitle


\section{Introduction}

Protein-based materials profit from the extensive tunability and combinatorial
diversity of their modular building blocks to create new and versatile
functionalities.\cite{VilenchikGriffithSt.ClairEtAl1998,MargolinNavia2001,ZhangWuYinEtAl2017,SleyterSchusterPum2003}
For instance, Suzuki and collaborators synthesized the
first Protein Crystals (PC) with auxetic behavior,\cite{SuzukiCardoneRestrepoEtAl2016}
\emph{i.e.}, stretching (shrinking) them along one axis results in
their expansion (compression) along a perpendicular one. The unusual
features displayed by auxetic materials make them appealing for many applications, \emph{e.g.},
in personal protective clothing,\cite{AldersonSimkins2005} clinical prosthesis,
\cite{Scarpa2008} filtration mechanisms,\cite{YangLiShiEtAl2004} mechanical
lungs,\cite{AldersonSimkins2005} controlled release of drugs,\cite{AldersonSimkins2005}
and reinforcement of composite materials.\cite{AldersonAlderson2007} Moreover, the auxetic PCs prepared by Suzuki and collaborators display coherent dynamics, as suggested
by Transmission Electron Microscopy studies, which reveal a continuous transition from an open (porous)
to a closed (tight-packed) configuration.\cite{SuzukiCardoneRestrepoEtAl2016} Unexpectedly, the geometrical rearrangements associated to this transition span across entire single crystals,
without identifiable formation of local configurational domains. Although the geometrical
arrangement in this system is closely related to that of the simple model
of rotating squares studied by Grima,\cite{GrimaEvans2000} the inclusion
of finite length linkages between the building blocks introduces further
degrees of freedom that increase the complexity of the system. Hence
the persistence of auxeticity or coherence is no longer an obvious feature of these structures.

From the the theoretical standpoint, it is appealing to explore the extent of inter-subunit coherent dynamics throughout the network, and the conditions leading to it. This exploration can be addressed with a computational approach with the aid of Molecular Dynamics
(MD) simulations. An all-atom simulation with four subunits has been already carried out to explore the thermodynamics and system-solvent interactions.\cite{AlbersteinSuzukiPaesaniEtAl2018} However, this approach proved to be extremely expensive from a computational standpoint, such that it only allowed to study the minimal non-trivial number of subunits. Nevertheless, upon examination, the geometry of the system suggests the sufficiency of a coarse-grained
framework in which the protein subunits may be suitably represented
by rigid squares whose generalized coordinates afford the exploration of dynamic variables within the formalism of constrained Lagrangian mechanics. Several methods
have been developed to address problems akin to the present one,\cite{NielsenLopezSrinivasEtAl2004} including adaptive time-step,\cite{BarthLeimkuhlerReich1999}
explicit minimization of discretized action,\cite{LeyendeckerHartmannKoch2012}
penalty functions,\cite{LeyendeckerHartmannKoch2012} event-driven dynamics,\cite{PenaZonSchofieldEtAl2007}
and impulsive constraints,\cite{StrattHolmgrenChandler1981} to name
a few.

On the other hand, in most MD simulations, constraints are employed
to freeze out only high-frequency vibrational modes, such as hydrogen atom bond vibrations in solvents and macromolecules,
and are rarely applied to the primary degrees of freedom of the system.\cite{KalyaanamoorthyChen2014} Furthermore, this typically serves to merely allow for a slightly longer integration time-step (typically from 1 to 2 fs) and does not constitute a significant element of the simulation.\cite{VanGunsterenKarplus1982,GanesanCooteBarakat2017} Hence, working examples of highly-constrained MD simulations are fairly scarce relative to their unconstrained counterparts. The recent realization
of auxetic PCs together with an increased interest in soft mechanical
materials demands new MD methodologies that can address the simulation
of highly articulated structures involving large numbers of geometrical
constraints.\cite{PoursinaBhaleraoFloresEtAl2011} In these instances, fulfillment of the constraints is achieved
by correcting an unconstrained update of the configuration at each
time-step.\cite{AllenTildesley2012} Currently, the most widespread
methods perform such corrections iteratively,\cite{RyckaertCiccottiBerendsen1977,Andersen1983}
which can be, from a computational standpoint, a shortcoming for highly constrained problems.\cite{KraeutlerGunsterenHuenenberger2001,NguyenPhillipsAndersonEtAl2011} There are alternative methodologies that address
these concerns\cite{Kneller2017,Garcia-RisuenoEcheniqueAlonso2011} by eliminating the iterations
but are prone to drift, thus requiring adaptive time-step schemes\cite{BarthLeimkuhlerReich1999}
which can become quite costly computationally. 

In this work, we investigate the coherent behavior by performing
a statistical analysis of trajectories simulated through a performance enhanced MD scheme.
We demonstrate that corrections to the unconstrained velocities
do not need to be calculated iteratively if the effects of the constraints
are regarded as impulses,\cite{StrattHolmgrenChandler1981} as it
is commonly done for collision responses.\cite{Coutinho2012}
Moreover, we introduce a novel method that corrects configurations to fulfill holonomic constraints in coordinate space, which is non-iterative and accurate enough to be implemented when the simulation requires the evaluation of shorter time-steps.
This new machinery, along with slight improvements to tools from the methods mentioned above, produces a significantly optimized approach that is able to deal with complex tasks such as the handling of collision events. The simulated trajectories produce a set of time series that are further analyzed to determine the degree of uniformity among the motion of the network components.

This manuscript is organized as follows: in Section \ref{sec:model}, we  introduce the coarse-grained model and identify the essential degrees of freedom
required to emulate the dynamics of the auxetic PC by explicit simulation. In Section \ref{sec:dynamics}, we describe the formalism of constrained
Lagrangian dynamics, which gives rise to the corresponding Equations Of
Motion (EOM). In Section \ref{sec:grima}, we illustrate the features of the system for conditions in which the EOM can be expressed analytically. In Section \ref{sec:numerical}, we introduce the numerical and computational methods developed
to handle the EOM integration, and the collision detection
and response protocols. In Section \ref{sec:2x2}, we study the smallest network and assess the stability of coherent dynamics that this structure might display. In Section \ref{sec:10x10}, we examine the dynamic behavior of a larger array under varying conditions, and contrast the relevance of the geometry with that of additional forces driving coherence. Finally, we present a
summary and conclusions of the work in Section \ref{sec:conclusion}.

\section{Definition of the coarse-grained model\label{sec:model}}

\begin{figure}[ht]
\protect\centering{}\protect\protect\includegraphics[scale=0.35]{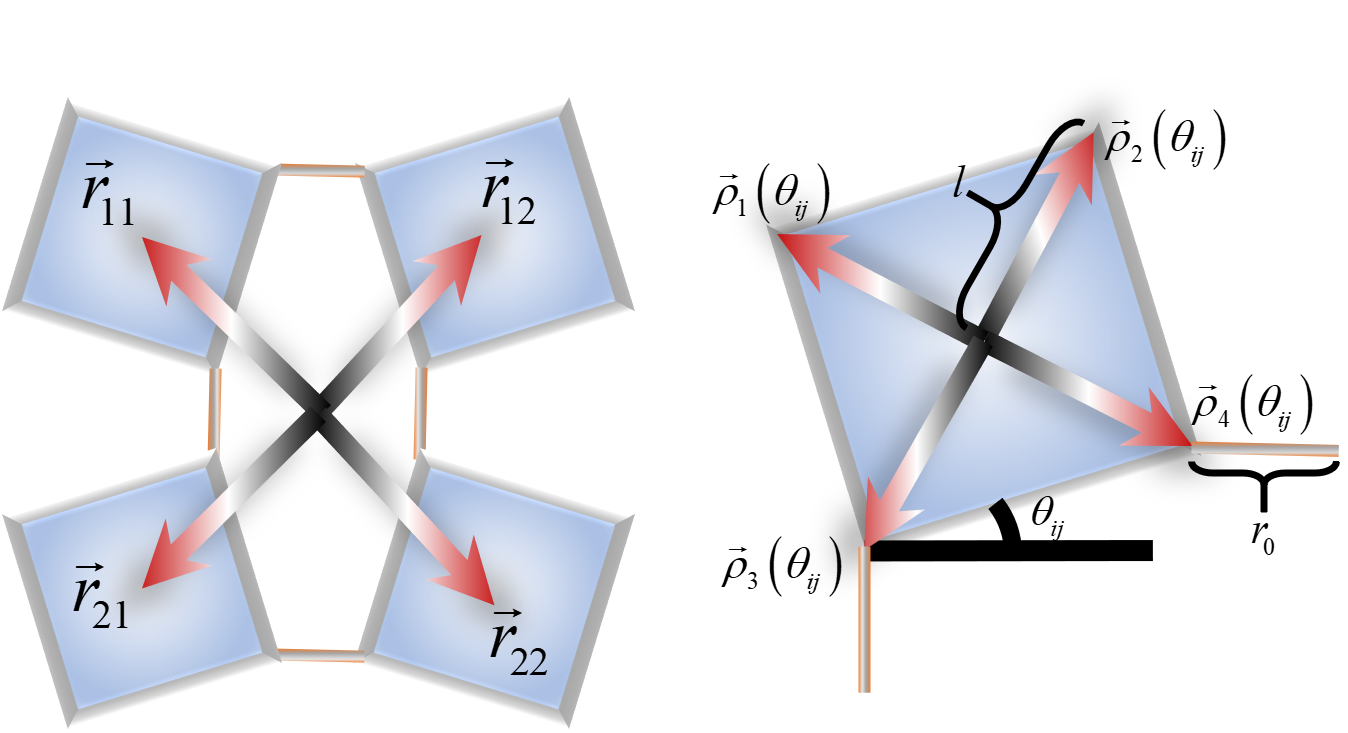}\protect\protect\caption{Elements and labels for the description of the network. Notice that for the open configuration, $|\theta_{ij}|=\pi/4$, while for the closed one, $|\theta_{ij}|=0$ or $\pi/2$.\label{fig:labang}}
\protect 
\end{figure}

The PC synthesized by Suzuki and co-workers was prepared through the self-assembly of \emph{L}-rhamnulose-1-phosphate aldolase units containing surface-exposed cysteines (\textsuperscript{C98}RhuA), which form intermolecular disulfide bonds to yield extended two-dimensional crystals.\cite{SuzukiCardoneRestrepoEtAl2016} Upon tessellation, the protein assemblage adopts a checkerboard pattern (\emph{p}4212 symmetry\cite{BurnsGlazerBurnsEtAl2013}) where each of the $C_{4}$-symmetric \textsuperscript{C98}RhuA subunits is linked to its four nearest neighbors through its vertices (see Fig. \ref{fig:RhuA}).\cite{SuzukiCardoneRestrepoEtAl2016} In our coarse-grained model, we consider the aforementioned articulated structure as a collection of rigid impenetrable squares connected by massless rigid rods. Labelling each square with indices \emph{ij} denoting its Cartesian position in the grid (Figure \ref{fig:labang}), its evolution is described by a 3D state vector $\vec{s}_{ij}$ and its time derivatives:

\begin{figure}[ht]
\protect\centering{}\protect\protect\includegraphics[scale=0.25]{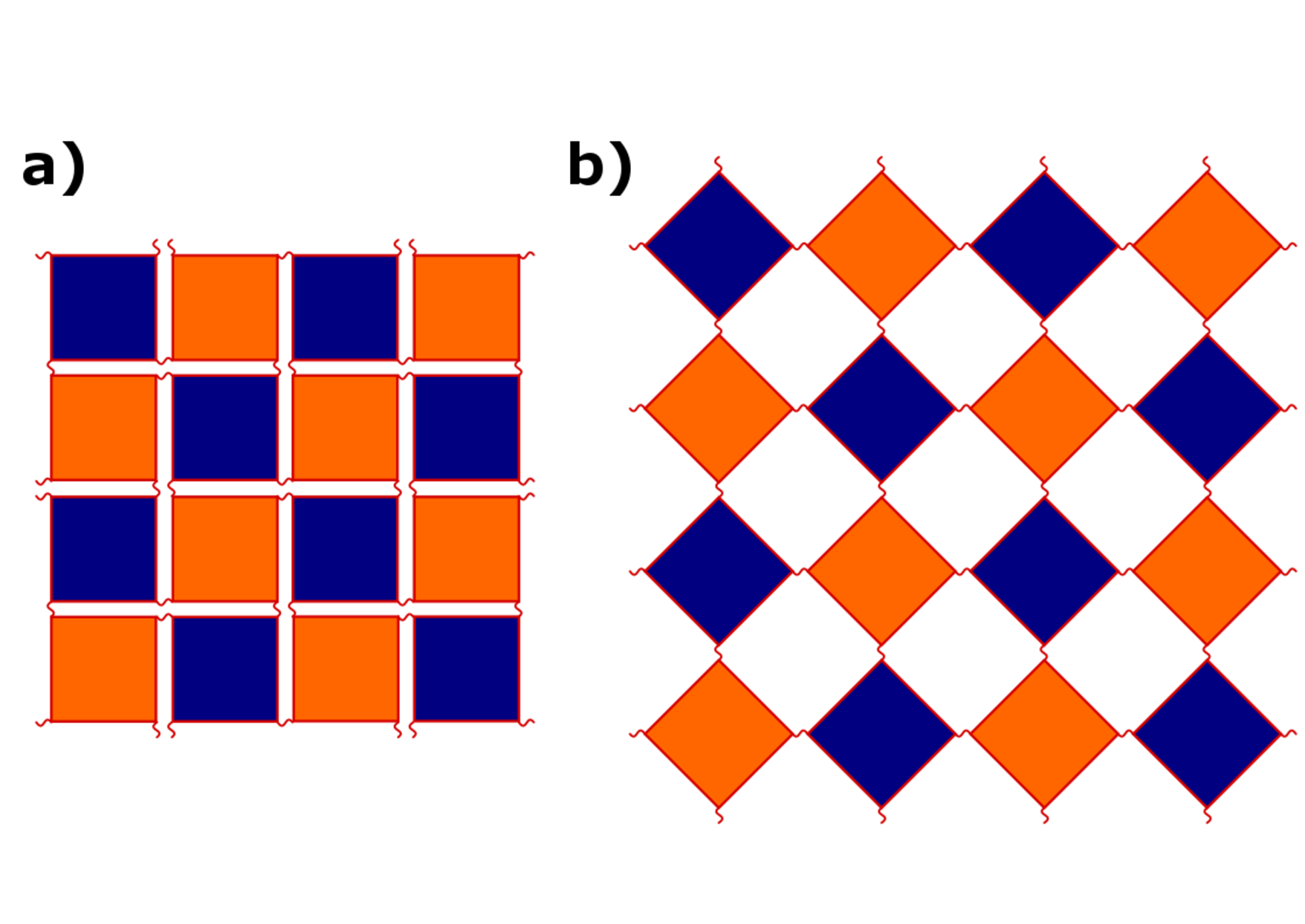}\protect\protect\caption{Closed (a) and open (b) configurations of \textsuperscript{C98}RhuA\label{fig:RhuA}. Curved lines represent linkages between protein units, and are regarded as rigid rods in our simulation. Protein units are colored blue and orange, denoting different faces of the protein. In our simulation, these differences are unimportant and therefore ignored. The protein units are simulated as rigid impenetrable squares. }
\protect 
\end{figure}

\begin{subequations}
\begin{align}
\vec{s}_{ij} &= \vec{r}_{ij}+\theta_{ij}\hat{e}_{z},\\
\dot{\vec{s}}_{ij} &= \vec{v}_{ij}+\omega_{ij}\hat{e}_{z},\\
\ddot{\vec{s}}_{ij} &= \vec{a}_{ij}+\alpha_{ij}\hat{e}_{z},
\end{align}
\end{subequations}
where $\vec{r}_{ij}$, $\vec{v}_{ij}=\dot{\vec{r}}_{ij}$ and $\vec{a}_{ij}=\ddot{\vec{v}}_{ij}$ denote
the location, velocity, and linear acceleration, respectively, of the
centroid of the square in two-dimensional Euclidean space; $\theta_{ij}$
is the orientation of the square with respect to a fixed lab frame, while
$\omega_{ij}=\dot{\theta}_{ij}$ and $\alpha_{ij}=\ddot{\theta}_{ij}$ are its angular velocity and angular
acceleration, respectively.

The vertices of the squares play a fundamental role in the dynamics of the
system. The two-dimensional position of the \emph{k}-th vertex belonging
to the \emph{ij}-th square (see Fig. \ref{fig:labang}) is given by

\begin{equation}
\vec{r}_{ijk}=\vec{r}_{ij}+\vec{\rho}_{k}(\theta_{ij}),
\end{equation}
where 
\begin{multline}
\vec{\rho}_{k}(\theta_{ij})=\frac{l}{\sqrt{2}}\left[\cos(k\pi)\mathbf{1}_{2}+\left(\sin\tfrac{k\pi}{2}-\cos\tfrac{k\pi}{2}\right)\mathbf{G}\right]\\
\cdot\left[\cos(\theta_{ij})\hat{e}_{x}+\sin(\theta_{ij})\hat{e}_{y}\right],
\end{multline}
is the vector pointing from the center of mass  to the $k$-th vertex in the $ij$-th subunit, \emph{l} is the half-diagonal length of the square, $\mathbf{1}_{2}=\hat{e}_{x}\hat{e}_{x}^{T}+\hat{e}_{y}\hat{e}_{y}^{T}$
is the two-dimensional projection operator, and $\mathbf{G}=\hat{e}_{y}\hat{e}_{x}^{T}-\hat{e}_{x}\hat{e}_{y}^{T}+\hat{e}_{z}\hat{e}_{z}^{T}$
is the \emph{SO}(2) generator of rotations.
The velocity of each vertex, $\vec{v}_{ijk}=\dot{\vec{r}}_{ijk}$, is calculated using that of the center
of mass, as well as the tangential velocity of the corresponding subunit:
\begin{equation}\label{eq:vertvel}
\vec{v}_{ijk}=\vec{v}_{ij}+\omega_{ij}\hat{e}_{z}\times\vec{\rho}_{k}(\theta_{ij}).
\end{equation}
Defining the additional projector 
\begin{equation}
\mathbf{B}_{ijk}=\mathbf{1}_{2}+\mathbf{G}\vec{\rho}_{k}(\theta_{ij})\hat{e}_{z}^{T},\label{eq:projvel}
\end{equation}
Eq. \eqref{eq:vertvel} can be written in terms
of the velocity vector of the center of mass of each square,
\begin{equation}
\vec{v}_{ijk}=\mathbf{B}_{ijk}\dot{\vec{s}}_{ij}.
\end{equation}
On the other hand, the acceleration of a given vertex must account for
the center of mass, radial (centripetal), and tangential accelerations,
\begin{equation}\label{eq:vertaccel}
\vec{a}_{ijk}=\mathbf{B}_{ijk}\ddot{\vec{s}}_{ij}-\omega_{ij}^{2}\vec{\rho}_{k}(\theta_{ij}).
\end{equation}
A given vertex \emph{k} in a protein unit \emph{ij} is linked to only one vertex \emph{k'} in another protein unit \emph{i'j'} (Fig. \ref{fig:labang}). The relations between the primed and unprimed indices are summarized in
Table \ref{tab:tabijk}.

\begin{table}
\caption{Relations between the  indices of a given vertex, \emph{ijk}, and those of the vertex  linked to it, \emph{i'j'k'}. Notice that these relations are self-inverse, \emph{i. e.}, $(i')'(j')'(k')'=ijk$.}\label{tab:tabijk}
\begin{ruledtabular}
\begin{tabular}{*{7}{c}}
\multirow{2}{*}{$k$ }  & \multicolumn{3}{c}{even $\left(i+j\right)$} & \multicolumn{3}{c}{odd $\left(i+j\right)$}\tabularnewline
 & $i'$  & $j'$  & $k'$  & $i'$  & $j'$  & $k'$\\
\hline 
1  & $i$  & $j-1$  & 2  & $i-1$  & $j$  & 3\\
2  & $i-1$  & $j$  & 4  & $i$  & $j+1$  & 1\\
3  & $i+1$  & $j$  & 1  & $i$  & $j-1$  & 4\\
4  & $i$  & $j+1$  & 3  & $i+1$  & $j$  & 2
\end{tabular}
\end{ruledtabular}
\end{table}

For an $n_{s}\times n_{s}$ lattice with open-boundary conditions,
there will be $n_{b}=2n_{s}\left(n_{s}-1\right)$ rigid rods (disulfide
bonds) connecting the square (protein) units. With the notation developed
so far, the vector along the linkage between the \emph{ijk}-th and \emph{i'j'k'}-th  vertices can be written as 
\begin{equation}
\Delta\vec{r}_{ijk}=\vec{r}_{ijk}-\vec{r}_{i'j'k'}.
\end{equation}
The constraints due to the rigidity of the rods can be expressed
as
\begin{equation}
\lvert\Delta\vec{r}_{ijk}\rvert^{2}=r_{0}^{2},\label{eq:qconstr}
\end{equation}
where $r_{0}\approx(3/22)l$ is the equilibrium length of the disulfide
bonds.\cite{SuzukiCardoneRestrepoEtAl2016} Differentiation of Eq. \eqref{eq:qconstr}
with respect to time yields a velocity 
\begin{equation}
\Delta\vec{r}_{ijk}^{T}\Delta\vec{v}_{ijk}=0,\label{eq:velcons}
\end{equation}
and acceleration constraints \cite{BarthLeimkuhlerReich1999}
\begin{equation}
\Delta\vec{r}_{ijk}^{T}\Delta\vec{a}_{ijk}+\lvert\Delta\vec{v}_{ijk}\rvert^{2}=0.\label{eq:accelcons}
\end{equation}

\begin{figure}[ht]
\protect\centering\protect\protect\includegraphics[scale=0.5]{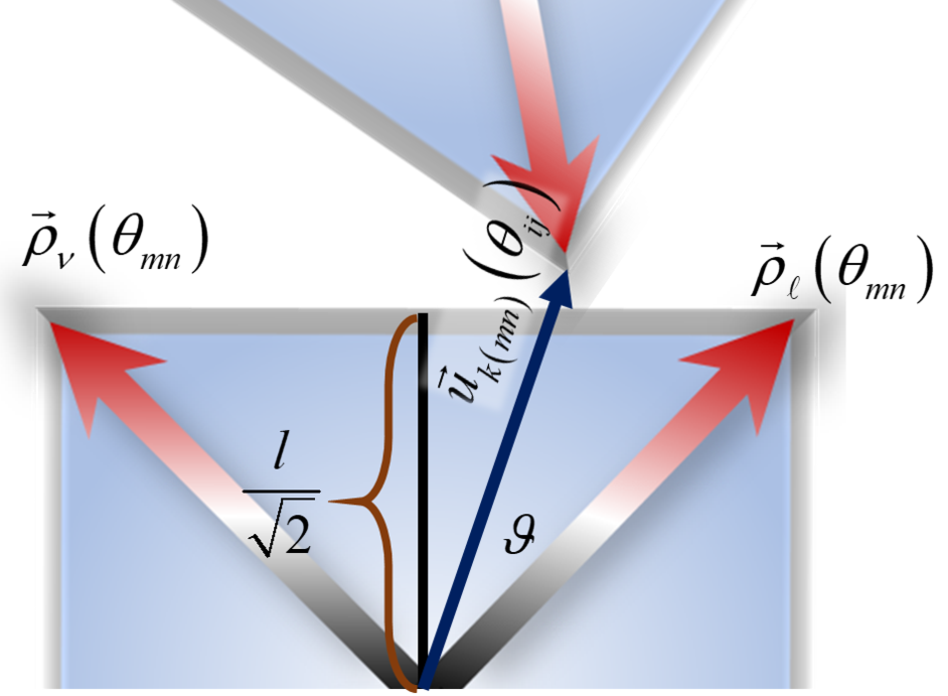}\protect\protect\caption{A pair of non-overlapping squares. The striker has not penetrated the area of the target. Hence, the projection of $\vec{u}_{k(mn)}(\theta_{ij})$ on the sides of the target is longer than the apothem. \label{fig:collis}}
\protect 
\end{figure}

Impenetrability is imposed by demanding that no vertex of any square shall be found inside the area of another. To formalize this demand in terms of the previously defined quantities, the \emph{ij}-th square is identified as a \emph{striker}, and the
\emph{mn}-th square as a \emph{target}.\cite{DonahueHrenyaZelinskayaEtAl2008} The squares do not overlap as long as, for all $k$-th vertices of the striker, the projections of the vector $\vec{u}_{k(mn)}(\theta_{ij})=\vec{r}_{ijk}-\vec{r}_{mn}$ on all the sides of the target remain longer than its apothem, $l/\sqrt{2}$. In Fig. \ref{fig:collis}, it can be seen that this projection can be written as
\begin{multline}
\frac{Z_{ijk}^{mn\ell}}{\sqrt{2}l}=\cos\left(\frac{\pi}{4}-\vartheta\right)\\
=\frac{1}{\sqrt{2}l}\left[\vec{\rho}_{\ell}(\theta_{mn})^{T}\vec{u}_{k(mn)}(\theta_{ij})\right.\\
\left.+\lvert\vec{\rho}_{\ell}(\theta_{mn})^{T}\mathbf{G}^{T}\vec{u}_{k(mn)}(\theta_{ij})\rvert\right],
\end{multline}
where $\ell$ labels the vertex of the target that is the closest to the $k$-th vertex of the striker, and $\vartheta$ is the angle between $\vec{u}_{k(mn)}(\theta_{ij})$ and $\vec{\rho}_\ell(\theta_{mn})$.
Consequently, the excluded volume constraint can be conveniently expressed as
\begin{equation}
Z^{mn\ell}_{ijk}-l^2\geq0,\label{eq:test}.
\end{equation}

Having settled the notation and key features of the
model, we proceed to discuss its dynamical evolution.

\section{Formulation of the dynamics\label{sec:dynamics}}

Since this work focuses mainly on the geometrical features of the
system, we do not take into account dissipative mechanisms; nevertheless, these and any additional potentials can be readily included in
generalizations of computational schemes compatible with the suggested
below.\cite{JohnsonLeyendeckerOrtiz2014} The system of concern is described
by a Lagrangian function per mass unit of the form
\begin{equation}
{\cal L}=\sum_{ij}\left(K_{ij}+V_{ij}+\sigma_{ij}+\sum_{mn}\varsigma_{ij(mn)}\right)\label{eq:lagrangian},
\end{equation}
where
\begin{equation}
K_{ij}=\frac{1}{2}\dot{\vec{s}}_{ij}^{T}\left(\mathbf{1}_{2}+I\hat{e}_{z}\hat{e}_{z}^{T}\right)\dot{\vec{s}}_{ij},
\end{equation}
is the kinetic energy with moment inertia $I=l^{2}/3$ for every square, and
\begin{equation}
V_{ij}=\frac{\kappa}{2}\left(\theta_{ij}^2-\langle\theta\rangle_0^2-\langle\theta\rangle_1^2\right)
\end{equation}
is a coherence inducing potential with intensity parameter $\kappa$,  orientation of reference $\langle\theta\rangle_0=n_s^{-1}\sum_{ij}\theta_{ij}$ (so that the orientation of the entire crystal is fixed), and an equilibrium orientation $\langle\theta\rangle_1=n_s^{-1}\sum_{ij}(-1)^{i+j}\theta_{ij}$ (which induces coherent behavior). The 
functions $\sigma_{ij}$ and $\varsigma_{ij(mn)}$ are, respectively,
holonomic and nonholonomic constraints\cite{Rosenberg1977}
standing for the rigid-rod linkages and the impenetrability of the squares.

\subsection{Holonomic constraints}

Constraints due to Eq. \eqref{eq:qconstr} are introduced into the
Lagrangian in the form: 
\begin{equation}
\sigma_{ij}=\sum_{k=1}^{4}\lambda_{ijk}\left(\lvert\Delta\vec{r}_{ijk}\rvert^{2}-r_{0}^{2}\right)=0,\label{eq:lconstr}
\end{equation}
where $\lambda_{ijk}$ are Lagrange multipliers characterizing the
rigid-rod linkages. Including only these constraints for the time
being, evaluation of the Euler-Lagrange equations with the Lagrangian in Eq. \eqref{eq:lagrangian} yields
\begin{equation}\label{eq:efforce}
\ddot{\vec{s}}_{ij}=f_{ij}\hat{e}_z+ \sum_{k=1}^{4}\lambda_{ijk}\mathbf{C}_{ijk}\Delta\vec{r}_{ijk},
\end{equation}
where $f_{ij}=-\kappa\left[\theta_{ij}-\langle\theta\rangle_0-(-1)^{i+j}\langle\theta\rangle_1\right]$, and the operator
\begin{equation}\label{eq:cproj}
\mathbf{C}_{ijk}=\mathbf{1}_{2}+\frac{\hat{e}_{z}\vec{\rho}_{k}(\theta_{ij})^{T}\mathbf{G}^{T}}{I}
\end{equation}
projects the vector $\Delta\vec{r}_{ijk}$
onto the acceleration $\vec{a}_{ijk}$ in Eq. \eqref{eq:vertaccel}.
Here, it becomes evident that the Lagrange multipliers $\lambda_{ijk}$
play the role of adaptive stiffness constants. To calculate the multipliers,
the vertex accelerations can be written in terms of the bond vectors
$\Delta\vec{r}_{ijh}$,\cite{Coutinho2012} 
\begin{equation}\label{eq:accelvlamb}
\vec{a}_{ijk}= \sum_{h=1}^{4}\lambda_{ijh}\mathbf{A}_{hk}^{(ij)}\Delta\vec{r}_{ijh}+\left(f_{ij}\mathbf{G}-\omega_{ij}^{2}\mathbf{1}_2\right)\vec{\rho}_{k}(\theta_{ij}),
\end{equation}
where $\mathbf{A}_{hk}^{(ij)}=\mathbf{B}_{ijk}\mathbf{C}_{ijh}$, and then plugged into the acceleration constraint of Eq. \eqref{eq:accelcons}:
\begin{multline}\label{eq:accelam}
-\lvert\Delta\vec{v}_{ijk}\rvert^{2}=\Delta\vec{r}_{ijk}^{T}\left[\sum_{h=1}^{4}\left(\lambda_{ijh}\mathbf{A}_{hk}^{(ij)}\Delta\vec{r}_{ijh}-\lambda_{i'j'h}\mathbf{A}_{h k'}^{(i'j')}\Delta\vec{r}_{i'j'h}\right)\right.\\
\left.+\left(\omega_{i'j'}^{2}\mathbf{1}_2-f_{i'j'}\mathbf{G}\right)\vec{\rho}_{k'}(\theta_{i'j'})+\left(f_{ij}\mathbf{G}-\omega_{ij}^{2}\mathbf{1}_2\right)\vec{\rho}_{k}(\theta_{ij})\right.\bigg].
\end{multline}

By defining the $n_{b}$-dimensional vectors whose $ijk$-th component is given by
\begin{widetext}
\begin{subequations}
\begin{align}\label{eq:vecdyn}
(\mathbf{b})_{ijk} &= \Delta\vec{r}_{ijk}^{T}\left[\left(\omega_{ij}^{2}\mathbf{1}_2-f_{ij}\mathbf{G}\right)\vec{\rho}_{k}(\theta_{ij})+\left(f_{i'j'}\mathbf{G}-\omega_{i'j'}^{2}\mathbf{1}_2\right)\vec{\rho}_{k'}(\theta_{i'j'})\right],\\
(\mathbf{w})_{ijk} &= \lvert\Delta\vec{v}_{ijk}\rvert^{2},\\
(\boldsymbol{\Lambda})_{ijk} &= \lambda_{ijk},\\
(\mathbf{M}\boldsymbol{\Lambda})_{ijk} &=\Delta\vec{r}_{ijk}^{T}\left[\sum_{h=1}^{4}\lambda_{ijh}\mathbf{A}_{hk}^{(ij)}\Delta\vec{r}_{ijh}-\sum_{\ell=1}^{4}\lambda_{i'j'\ell}\mathbf{A}_{\ell k'}^{(i'j')}\Delta\vec{r}_{i'j'\ell}\right],\label{eq:mlamb}
\end{align}
\end{subequations}
\end{widetext}
Eq. \eqref{eq:accelam} can be written in the form
\begin{equation}
\mathbf{M}\boldsymbol{\Lambda=b-w},\label{eq:lambex}
\end{equation}
which allows for the solution of the Lagrange multipliers upon inversion
of the matrix $\mathbf{M}$.\cite{Kneller2017} While, for the sake of brevity, we are not providing the explicit form of the matrix elements, the form of $\mathbf{M}$ can be readily extracted from Eq. \eqref{eq:mlamb}.

\subsection{Nonholonomic constraints}

Inequality \eqref{eq:test} leads to a set of constraints for
the allowed (\emph{i.e.}, non-interpenetrated) configurations of the
system,

\begin{equation}
\label{eq:kktcons}\varsigma_{ij(mn)}=\sum_{k,\ell}\lambda_{ij(mn)}\Theta(l^2-Z^{mn\ell}_{ijk})=0,
\end{equation}
where $\Theta(x)$ is the Heaviside step function, and the coefficients $\lambda_{ij(mn)}$ are known as Karush-Kuhn-Tucker
(KKT) multipliers in the mathematical optimization literature.\cite{Kuhn2014,Karush2014,Gunaratne2006}
These constraints vanish whenever Eq. \eqref{eq:test} is a strict inequality,
thus justifying our disregard for them in the
previous section. With the same treatment as before, the inclusion
of these constraints in the EOM modifies the effective forces of Eq.
\eqref{eq:efforce} as

\begin{subequations}
\begin{align}\label{eq:accol}
\ddot{\vec{s}}_{ij} &= f_{ij}\hat{e}_z+\sum_{k=1}^{4}\lambda_{ijk}\mathbf{C}_{ijk}\Delta\vec{r}_{ijk}-\sum_{mn}\mathbf{C}_{ij(mn)}\vec{J}_{\text{col}}^{ij(mn)},\\
\ddot{\vec{s}}_{mn}&=f_{mn}\hat{e}_z+ \sum_{k=1}^{4}\lambda_{mnk}\mathbf{C}_{mnk}\Delta\vec{r}_{mnk}+\sum_{ij}\mathbf{C}_{mn(ij)}\vec{J}_{\text{col}}^{ij(mn)},
\end{align}
\end{subequations}
where 
\begin{equation}
\vec{J}_{\text{col}}^{ij(mn)}=\lambda_{ij(mn)}\frac{\vec{\rho}_{\ell}(\theta_{mn})+\vec{\rho}_{\nu}(\theta_{mn})}{\sqrt{2}l}\delta(l^2-Z_{ijk}^{mn\ell})\label{eq:impulse}
\end{equation}
is an impulsive force accounting for the collisions, with $\delta(x)$ the Dirac delta function, $\nu$ labeling
the second vertex in the target closest to the striker vertex, and
the projection operators 
\begin{subequations}\label{eq:projkkt}
\begin{align}
\mathbf{C}_{ij(mn)}&= \mathbf{1}_{2}+\frac{\hat{e}_{z}\vec{\rho}_{k(mn)}(\theta_{ij})^{T}\mathbf{G}^{T}}{I},\\
\mathbf{C}_{mn(ij)}&= \mathbf{1}_{2}+\frac{\hat{e}_{z}\vec{u}_{k(mn)}(\theta_{ij})^{T}\mathbf{G}^{T}}{I},
\end{align}
\end{subequations}
having analogous interpretations as those defined in Eq. \eqref{eq:cproj}, where $\vec{\rho}_{k(mn)}(\theta_{ij})$ locates the striker vertex in the frame of its own subunit. It is worthwhile noticing that, while they can be formally derived through meticulous rigid-body analysis as in Ref. \onlinecite{Coutinho2012}, the impulsive force in Eq. \eqref{eq:impulse}, and its projectors in Eq. \eqref{eq:projkkt}, are readily obtained from  the Euler-Lagrange equations with the constraint in Eq. \eqref{eq:test}.

\section{Analytical example: Grima's model \label{sec:grima}}

As a reference, we show the behavior of the system in the case in which absolute coherence is among its inherent features.\cite{GrimaEvans2000} To impose uniform motion, we add the holonomic constraints
\begin{equation}\label{eq:angconstr}
\theta_{ij}=-\theta_{i,j+1}=-\theta_{i+1,j}
\end{equation}
to the Lagrangian in Eq. \eqref{eq:lagrangian}. Under such conditions, $\sum_{ij}V_{ij}=0$ regardless of the value of $\kappa$.

 It is worthwhile noticing that the constraints in Eq. \eqref{eq:angconstr} emerge naturally from Eq. \eqref{eq:qconstr} when $r_0=0$, which corresponds exactly to Grima's model. For structures with linkages of non-zero length, the set of all holonomic constraints introduces periodicity as well as some redundancies. After removing the latter, it becomes clear that the only remaining degrees of freedom of the system are $\langle\theta\rangle_1$ and its time derivative, $\langle\omega\rangle_1$. Taking advantage of the simplified structure of the problem, we arrive analytically to the kinematic EOM:
\begin{widetext}
\begin{subequations}\label{eq:thetat}
\begin{align}
E\left[\tfrac{\pi}{4}-\langle\theta(t)\rangle_1,2(1-n_s^2)\right]=&E\left[\tfrac{\pi}{4}-\langle\theta(0)\rangle_1,2(1-n_s^2)\right]-\sqrt{n_s^2+(1-n_s^2)\sin\left[2\langle\theta(0)\rangle_1\right]}\langle\omega(0)\rangle_1 t,\\
\left\{n_s^2+(1-n_s^2)\sin\left[2\langle\theta(t)\rangle_1\right]\right\}\langle\omega(t)\rangle_1^2=&\left\{n_s^2+(1-n_s^2)\sin\left[2\langle\theta(0)\rangle_1\right]\right\}\langle\omega(0)\rangle_1^2,\\
2l\sin(2\langle\theta(t)\rangle_1)+r_0\sin\left(\langle\theta(t)\rangle_1+\frac{\pi}{4}\right)\geq&0,\label{eq:grimmaconst}
\end{align}
\end{subequations}
\end{widetext}
where $E(\varphi,m)=\int_0^\varphi\sqrt{1-m\sin^2\phi}\,d\phi$ is the incomplete elliptic integral of the second kind,\cite{AbramowitzStegun1965}, and $\langle\omega(t)\rangle_1=\langle\dot\theta(t)\rangle_1$. The non-holonomic constraint in Eq. \eqref{eq:grimmaconst} originates from the combination of all the constraints in the model.

\begin{figure}[ht]
\protect\centering\protect\protect\includegraphics[scale=.4]{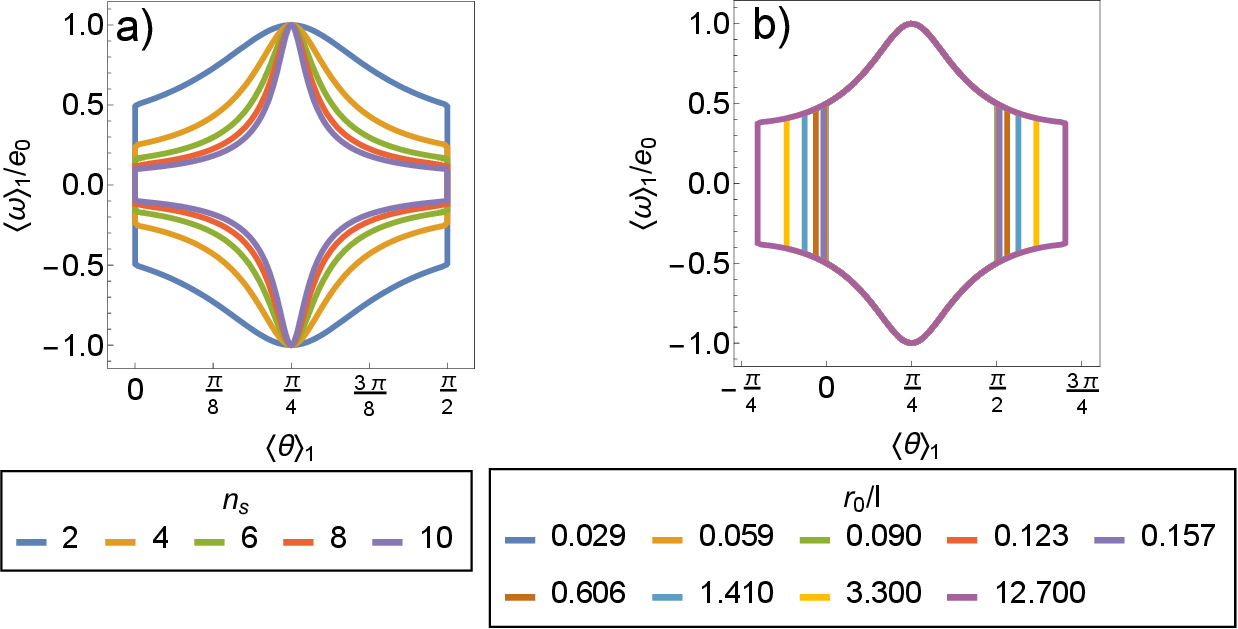}\protect\protect\caption{Phase space of the dynamics of Grima's model with $e_0=\langle\omega(0)\rangle_1\sqrt{n_s^2+(1-n_s^2)\sin\left[2\langle\theta(0)\rangle_1\right]}$. (a) Lattices of several sizes with $r_0=0$. As the lattice grows, the kinetic energy in the open configuration increases with respect to the closed ones. (b)  $2\times2$ lattice with several link-length-to-side ratios. The expanded domain of the angle variable reflects the fact that linkages enable a wider range of motion.\label{fig:phsapc}}
\protect 
\end{figure}

An analysis of the phase space of this system (Fig. \ref{fig:phsapc}) reveals that the angular velocity experiences sudden discontinuous sign switching without going smoothly through zero. This is the signature of the collision in the closed configuration that limits the span of the angle. In addition, the angular velocity has a maximum at the open configuration and a minimum at the closed one. The implication is that when $\langle\theta\rangle_1\approx\pi/4$ the squares have maximum freedom and the energy is mostly rotational. In contrast, for angles close to 0 or $\pi/2$ , the motion of the lattice is predominantly translational since the squares are experiencing the required displacement for an ordered packing. This interpretation is consistent with the fact that larger lattices yield more kinetic energy to translations near the closed configuration as evidenced in Fig. \ref{fig:phsapc}a. 

\section{Numerical simulations\label{sec:numerical}}

The dynamics of {\textbf{s}} (the set of all state vectors $\vec{s}_{ij}$) is propagated by means of a Störmer-Verlet (SV) time integration\cite{SwopeAndersenBerensEtAl1982}:
\begin{subequations}
\begin{align}
\mathbf{s}(t+\Delta t)&= \mathbf{s}(t)+\dot{\mathbf{s}}(t)\Delta t+\frac{\ddot{\mathbf{s}}(t)}{2}\Delta t^{2},\label{eq:verletq}\\
\dot{\mathbf{s}}(t+\Delta t)&= \dot{\mathbf{s}}(t)+\frac{\ddot{\mathbf{s}}(t)+\ddot{\mathbf{s}}(t+\Delta t)}{2}\Delta t.\label{eq:verletp}
\end{align}
\end{subequations}


It is worth noting that the Lagrange multipliers computed with Eq.
\eqref{eq:lambex} are, in principle, exact.\cite{Kneller2017} However, time discretization introduces error that propagates during the dynamics, which results in drift. As a consequence, the constraints are not exactly
fulfilled throughout the dynamics.\cite{AllenTildesley2012} This
issue is solved by computing the Lagrange multipliers so that the
constraints are satisfied at each time step. Since there are constraints
in both coordinate and velocity spaces, there will be a set of Lagrange
multipliers, $\lambda^{(q)}$, obtained from subjecting the coordinates
in Eq. \eqref{eq:verletq} to the constraint given in Eq. \eqref{eq:qconstr},\cite{RyckaertCiccottiBerendsen1977}
in addition to a set, $\lambda^{(p)}$, due to the enforcement of Eq. \eqref{eq:velcons}
by  the velocities in Eq. \eqref{eq:verletp}.\cite{Andersen1983}
The strategies employed for each set are described below.

\subsection{Collision-free dynamics}

Numerical accuracy can be significantly improved by using a control function dependent on the Lagrange multipliers, which sets the size of the time-step in an adaptive framework.\cite{BarthLeimkuhlerReich1999} However, for the purposes of this work, we have found that a constant time-step with numerical value $\Delta t=0.01$ provides a fair balance between stability and implementation time. By equating the number of time-steps it takes to complete a transition between the open and the closed configuration in the present coarse-grained simulation to the time it takes for the same event to happen in network model simulations\cite{LymanPfaendtnerVoth2008,AlbersteinSuzukiPaesaniEtAl2018} for various system sizes, we have determined that our time-step size is of the order of a few nanoseconds.

\subsubsection{Coordinate constraints.}

Equation \eqref{eq:qconstr} together with the integration step of Eq.
\eqref{eq:verletq} give rise to the SHAKE equations for this system
\cite{RyckaertCiccottiBerendsen1977}: 
\begin{subequations}
\begin{align}
\begin{split}
\label{eq:shake}\vec{s}_{ij}(t+\Delta t)= &\vec{s}_{ij}(t)+\dot{\vec{s}}_{ij}(t)\Delta t\\
+\Bigg( f_{ij}(t)\hat{e}_z&\left.+\sum_{k=1}^{4}\lambda_{ijk}^{(q)}(t)\mathbf{C}_{ijk}(t)\Delta\vec{r}_{ijk}(t)\right)\frac{\Delta t^{2}}{2},
\end{split}\\
\lvert\Delta\vec{r}_{ijk}(t+\Delta t)\rvert^{2}=& r_{0}^{2},
\end{align}
\end{subequations}
where the unknowns are the updated coordinates and the Lagrange
multipliers. The quadratic character of the constraint, as well as
the trigonometric functions involved in the calculation of the inter-vertex
distances, make this scheme highly nonlinear. Fortunately, the values
of the Lagrange multipliers obtained with Eq. \eqref{eq:lambex},
although inaccurate for the integration scheme, usually require but small corrections. In this
work, such values are used as initial guesses for a Levenberg-Marquardt algorithm\cite{Levenberg1944,Marquardt1963}
carried out by the built-in MATLAB gradient-based nonlinear solver
\texttt{fsolve}.\cite{MathWorks2019}

\subsubsection{Velocity constraints.}

The velocity step of the time integration [Eq. \eqref{eq:verletp}]
together with the velocity constraint of Eq. \eqref{eq:velcons} form
the RATTLE equations of the system.\cite{Andersen1983} The latter
is a set of linear equations that can be written in matrix form as
\begin{equation}\label{eq:mtmbnd}
\begin{pmatrix}
\mathbf{1}_{3n_{s}^{2}} & \mathcal{C}\\
\mathcal{B} & \mathbf{0}_{n_{b}}
\end{pmatrix}
\begin{pmatrix}
\dot{\mathbf{s}}(t+\Delta t)\\
\boldsymbol{\Lambda}^{(p)}\frac{\Delta t}{2}
\end{pmatrix}
=\begin{pmatrix}
\dot{\mathbf{s}}(t)+\ddot{\mathbf{s}}(t)\frac{\Delta t}{2}\\
\mathbf{0}_{n_{b}\times1}
\end{pmatrix},
\end{equation}
where $\mathcal{C}$ is a $3n_{s}^{2}\times n_{b}$ matrix with blocks
$\mathrm{sgn}(i'j'-ij)\mathbf{C}_{ijk}(t+\Delta t)\Delta\vec{r}_{ijk}(t+\Delta t)\Delta t/2$,
and $\mathcal{B}$ is a $n_{b}\times3n_{s}^{2}$ matrix with blocks
given by $\mathrm{sgn}(ij-i'j')\Delta\vec{r}_{ijk}^{T}(t+\Delta t)\mathbf{B}_{ijk}(t+\Delta t)$.
Since the effects of the constraints in the EOM can be integrated as a set of effective impulsive
forces,\cite{StrattHolmgrenChandler1981} the leftmost matrix in
Eq. \ref{eq:mtmbnd} will be referred to as the Momentum Transfer
Matrix (MTM). From Eq. \ref{eq:mtmbnd}, it follows that the updated
velocities as well as the Lagrange multipliers can be computed from
the inversion of the MTM in a single step.

\subsection{Dynamics including collisions}

To detect whether inequality \eqref{eq:test} is fulfilled, the first step
is a pairwise pruning procedure that checks the distances between centers of mass
for each pair of squares. If this distance is larger than the diagonal of the squares, $2l$, the collision is rejected; otherwise, all the vertices
of the pair of squares being considered are tested in detail with
Eq. \eqref{eq:test}, trying and swapping the roles of striker and
target for both squares. If the configuration
$\mathbf{s}(t+\Delta t)$ contains violations of the inequality in Eq. \eqref{eq:test} during the dynamics,
the collisions are then accepted and the indices of both the target and the
vertex of the striker are stored to be later used in the collision
handling protocol. Having identified the overlapping squares and vertices, the next step
is to find the time at which the contact takes place as accurately as possible. Such task
would require solving the system of nonlinear Eqs. \eqref{eq:shake} multiple
times, which, given the iterative nature of the method hitherto employed,
might render it computationally expensive. On the other hand, for very small time-steps $\delta t\ll\Delta t$,
we have
\begin{equation}
\begin{pmatrix}\cos\left(\omega\delta t+\alpha\frac{\delta t^{2}}{2}\right)\\ \sin\left(\omega\delta t+\alpha\frac{\delta t^{2}}{2}\right)\end{pmatrix}
\approx
\begin{pmatrix}1\\\omega\delta t+\alpha\frac{\delta t^{2}}{2}\end{pmatrix}.
\end{equation}
This fact allows us to formulate a non-iterative method that improves upon the Lagrange multipliers calculated from Eq. \eqref{eq:lambex} by explicitly solving the coordinate constraints in Eq. \eqref{eq:qconstr}. More specifically, since the
bond lengths need to remain fixed, the EOM can be solved with the ansatz 
\begin{equation}
\begin{split}
\Delta\underline{\vec{r}}_{ijk}(t+\delta t)=&\Delta\vec{r}_{ijk}(t)\cos[\phi_{ijk}(t,\delta t)]\\
&+\frac{\Delta\vec{v}_{ijk}(t)}{\omega_{ijk}(t)}\sin[\phi_{ijk}(t,\delta t)]\label{eq:ansatz}
\end{split}
\end{equation}
where 
\begin{equation}
\omega_{ijk}(t)=\frac{\lvert\Delta\vec{v}_{ijk}(t)\rvert}{\lvert\Delta\vec{r}_{ijk}(t)\rvert}=\frac{\lvert\Delta\vec{v}_{ijk}(t)\rvert}{r_{0}}.
\end{equation}
It can be shown that (see Appendix \ref{sec:angal}), for an angle variable of the form $\phi_{ijk}(t,\delta t)=\omega_{ijk}(t)\delta t+\frac{\alpha_{ijk}(t)}{2}\delta t^{2}$,
the angular acceleration $\alpha_{ijk}(t)$ is
\begin{equation}
\alpha_{ijk}(t)=\frac{\lvert\Delta\vec{r}_{ijk}(t)^{T}\Delta\vec{a}_{ijk}(t)\rvert^{1/2}\Delta\vec{v}_{ijk}(t)^{T}\Delta\vec{a}_{ijk}(t)}{\lvert\Delta\vec{v}_{ijk}(t)\rvert^{2}r_{0}},
\end{equation}
where the acceleration differences $\Delta\vec a_{ijk}(t)$ can be found from Eq. \eqref{eq:accelcons}
using the Lagrange multipliers calculated with Eq. \eqref{eq:lambex}.
Since the vectors, $\Delta\underline{\vec{r}}_{ijk}$, in Eq. \eqref{eq:ansatz}
already fulfill the coordinate constraints in Eq. \eqref{eq:qconstr},
the latter can be replaced by 
\begin{equation}
\underline{\sigma}_{ijk}=\vec{\chi}_{ijk}^{T}\left(\Delta\vec{r}_{ijk}-\Delta\underline{\vec{r}}_{ijk}\right),
\end{equation}
where $\vec{\chi}_{ijk}$ is a vector of Lagrange multipliers. By
considering such constraints as the source of acceleration in the
integration step of Eq. \eqref{eq:verletq}, the Lagrange multipliers
can be found by solving the following system of  linear equations:
\begin{multline}\label{eq:acoord}
\sum_{h=1}^{4}\left(\mathbf{A}_{hk}^{(ij)}(t)\vec{\chi}_{ijh}^{(r)}(t)-\mathbf{A}_{h k'}^{(i'j')}(t)\vec{\chi}_{i'j'h}^{(r)}(t)\right) = (\mathbf{b})_{ijk}(t)\\
+\frac{2}{\delta t}\left(\frac{\Delta\underline{\vec{r}}_{ijk}(t+\delta t)-\Delta\vec{r}_{ijk}(t)}{\delta t}-\Delta\vec{v}_{ijk}(t)\right),
\end{multline}
and then used to update the coordinates with
\begin{equation}\label{eq:qcoord}
\begin{split}
\vec{s}_{ij}(t+\delta t)=& \vec{s}_{ij}(t)+\dot{\vec{s}}_{ij}(t)\delta t\\
&+\left(f_{ij}(t)\hat{e}_z+\sum_{k=1}^{4}\mathbf{C}_{ijk}(t)\vec{\chi}_{ijk}^{(r)}(t)\right)\frac{\delta t^{2}}{2}.
\end{split}
\end{equation}
The advantage of this method over the previous one is that it requires
solving Eqs. \eqref{eq:lambex}, \eqref{eq:acoord}, and \eqref{eq:qcoord}, which are all linear,  instead of Eq. \eqref{eq:shake}, which is a mixed system
of quadratic and trigonometric equations. Nevertheless, the
equivalence between the tangential behavior of the vertices and the
angular behavior of the squares assumed by this method is only valid
in the infinitesimal limit $\delta t\to0$. Therefore, this approach will be
exclusively used in the search for collision times, where the time-steps
are presumed to be small enough. Also, to reduce the truncation error
even more, we exploit the time-reversible character of the SV
integration by looking for a collision time starting from both $\mathbf{s}(t)$
and $\mathbf{s}(t+\Delta t)$, propagating the dynamics forward and backward
respectively, and keeping only the collision time, $\delta^{(\pm)}t_i$, that is the closest to the starting point. The collision times are found testing the inequality in Eq. \eqref{eq:test}
with the configurations computed with Eq. \eqref{eq:qcoord} via the built-in
MATLAB function \texttt{fzero}.\cite{MathWorks2019a}

Let $i$ be the $i$-th detected collision, the collision handling protocol thus becomes 
\begin{enumerate}
\item Starting from $\mathbf{s}(t)$, determine the forward time-step size
$0<\delta^{(+)}t_{i}<\Delta t$ for all $i$. 
\item Starting from $\mathbf{s}(t+\Delta t)$, determine the backward time-step
size $0>\delta^{(-)}t_{i}>-\Delta t$ for all $i$.
\item Calculate $\delta t_{i}=
\begin{cases}
\delta^{(+)}t_{i} & \text{if}\enskip\delta^{(+)}t_{i}<\lvert\delta^{(-)}t_{i}\rvert\\
\Delta t+\delta^{(-)}t_{i} & \text{if}\enskip\delta^{(+)}t_{i}>\lvert\delta^{(-)}t_{i}\rvert
\end{cases}
$ for all $i$.
\item Compute $\langle\delta t\rangle=\frac{1}{n_c}\sum_{i=1}^{n_c} \delta t_i$, where $n_c$ is the number of detected collisions. This choice introduces an error of roughly the same order of magnitude as the introduced by previous approximations, yet simplifies the execution of the protocol.
\item Evolve the system from $\left[\mathbf{s}(t),\dot{\mathbf{s}}(t)\right]$
to $\left[\mathbf{s}(t+\langle\delta t\rangle),\dot{\mathbf{s}}(t+\langle\delta t\rangle)\right]$
and perform collision response (Sec. \ref{sec:colresp}) to find $\left[{\mathbf{s}}^{\star}(t+\langle\delta t\rangle),{\dot{\mathbf{s}}}^{\star}(t+\langle\delta t\rangle)\right]$,
where the stars indicate ``after collision''. 
\item Evolve the system with a time-step of size $\Delta t-\langle\delta t\rangle$
to get to a corrected $\left[\mathbf{s}(t+\Delta t),\dot{\mathbf{s}}(t+\Delta t)\right]$. 
\item Return to collision-less dynamics. 
\end{enumerate}
\vfill
\subsubsection{Collision response.\label{sec:colresp}}

When two squares come into contact, the KKT multiplier associated
with the striker-target pair can have non-zero values. These
multipliers correspond to magnitudes of impulses which provide an instantaneous modification
to the velocities of the system. The effect of a bounce is the reflection
of the motion with respect to the plane normal to the collision, which
is consistent with the direction derived in Eq. \eqref{eq:impulse}.
This effect can be summarized in the following system
of equations: 
\begin{widetext}
\begin{subequations}
\begin{align}\label{eq:sincol}
{\dot{\vec{s}}}_{ij}^{\star}&= \dot{\vec{s}}_{ij}-\sum_{mn}\mathbf{C}_{ij(mn)}\vec{J}_{\text{col}}^{ij(mn)},\\
{\dot{\vec{s}}}_{mn}^{\star}&= \dot{\vec{s}}_{mn}+\sum_{mn}\mathbf{C}_{mn(ij)}\vec{J}_{\text{col}}^{ij(mn)},\\
\hat{n}_{ij(mn)}^{T}\left(\mathbf{B}_{ij(mn)}{\dot{\vec{s}}}_{ij}^{\star}-\mathbf{B}_{mn(ij)}{\dot{\vec{s}}}_{mn}^{\star}\right)&= \hat{n}_{ij(mn)}^{T}\left(\mathbf{B}_{mn(ij)}\dot{\vec{s}}_{mn}-\mathbf{B}_{ij(mn)}\dot{\vec{s}}_{ij}\right),
\end{align}
\end{subequations}
\end{widetext}
where the projection operators 
\begin{subequations}
\begin{align}
\mathbf{B}_{ij(mn)}&= \mathbf{1}_{2}+\mathbf{G}\vec{\rho}_{k(mn)}(\theta_{ij})\hat{e}_{z}^{T},\\
\mathbf{B}_{mn(ij)} &=\mathbf{1}_{2}+\mathbf{G}\vec{u}_{k(mn)}(\theta_{ij})\hat{e}_{z}^{T},
\end{align}
\end{subequations}
are analogous to those defined in Eq. \eqref{eq:projvel}. Nevertheless,
the impulse in Eq. \eqref{eq:impulse} can only describe vertex-side
collisions as the normal-to-collision plane is ill-defined for vertex-vertex
collisions (side-side collisions result from a pair of reciprocating
vertex-side collisions), in which case, solving Eq. \eqref{eq:sincol}
gives the vector 
\begin{multline}
\vec{n}_{\text{col}}^{ij(mn)}=\mathbf{G}\left(\frac{\vec{\rho}_{\ell}(\theta_{mn})\vec{\rho}_{\ell}(\theta_{mn})^{T}-\vec{\rho}_{k}(\theta_{ij})\vec{\rho}_{k}(\theta_{ij})^{T}}{I}\right)^{-1}\\
\cdot\left[\left(\mathbf{G}^{T}+\vec{\rho}_{\ell}(\theta_{mn})\hat{e}_{z}^{T}\right)\dot{\vec{s}}_{mn}+\left(\mathbf{G}^{T}+\vec{\rho}_{k}(\theta_{ij})\hat{e}_{z}^{T}\right)\dot{\vec{s}}_{ij}\right]
\end{multline}
which is parallel to the impulse. The structure of the system of equations
in Eq. \eqref{eq:sincol} allows for its seamless insertion into the
MTM such that, 
\begin{equation}\label{eq:colcons}
\begin{pmatrix}
\mathbf{1}_{3s^{2}} & \begin{array}{cc}
\mathcal{C}_{\text{bnd}} & \mathcal{C}_{\text{col}}\end{array}\\
\begin{array}{c}
\mathcal{B}_{\text{bnd}}\\
\mathcal{B}_{\text{col}}
\end{array} & \mathbf{0}_{3\left(n_{b}+n_{c}\right)\times\left(n_{b}+n_{c}\right)}
\end{pmatrix}
\begin{pmatrix}
\dot{\mathbf{s}}^{\star}\\
\boldsymbol{\Lambda}_{\text{bnd}}\\
\boldsymbol{\Lambda}_{\text{col}}
\end{pmatrix}
=\begin{pmatrix}
\dot{\mathbf{s}}\\
\mathbf{0}_{n_{b}\times1}\\
-\mathcal{B}_{\text{col}}\dot{\mathbf{s}}
\end{pmatrix}
\end{equation}
where the subscript 'bnd' has been added to differentiate whether
a quantity is related to holonomic (fixed bond lengths) or nonholonomic (excluded space) constraints.

\begin{figure}[ht]
\protect\centering{}\protect\protect\includegraphics[scale=0.24]{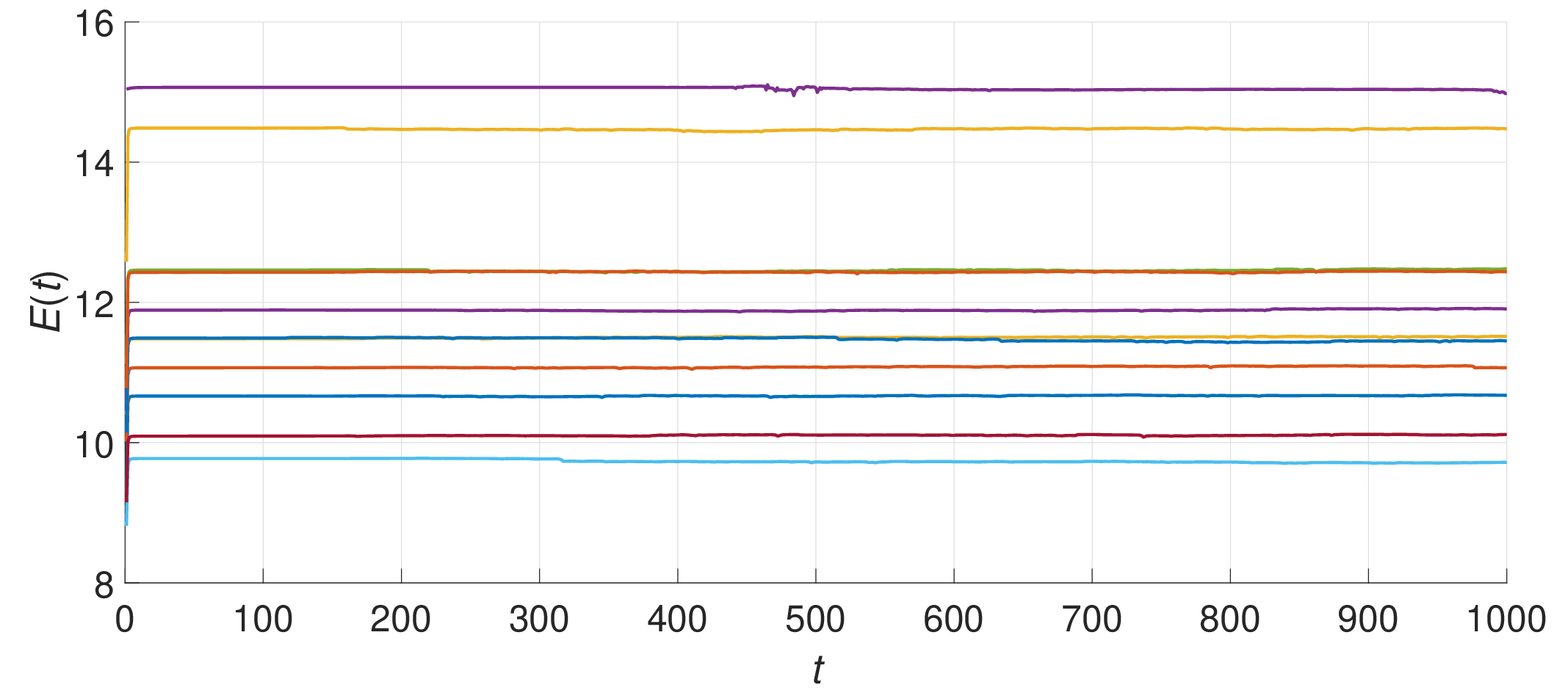}\protect\protect\caption{Energy evolution of trajectories with random initial conditions and $\kappa=0$ for a structure of size $10\times10$. \label{fig:enet}}
\protect 
\end{figure}

Figure \ref{fig:enet} shows the energy over time, $E(t)$, for trajectories of $10\times10$ arrays initialized over a range of random initial conditions. The observed stability of this quantity is indicative of the stability of the simulations.

\section{The 2$\times$2 structure\label{sec:2x2}}

The simplest array is that with four subunits, \emph{i.e.} a single pore. Alberstein and co-workers\cite{AlbersteinSuzukiPaesaniEtAl2018} present an analysis of its thermodynamic properties as obtained from an all-atom simulation. In the present work, we use the aforementioned techniques to describe its dynamics and introduce concepts that enable the description of larger structures.

\begin{figure}[ht]
\protect\centering{}\protect\protect\includegraphics[scale=0.4]{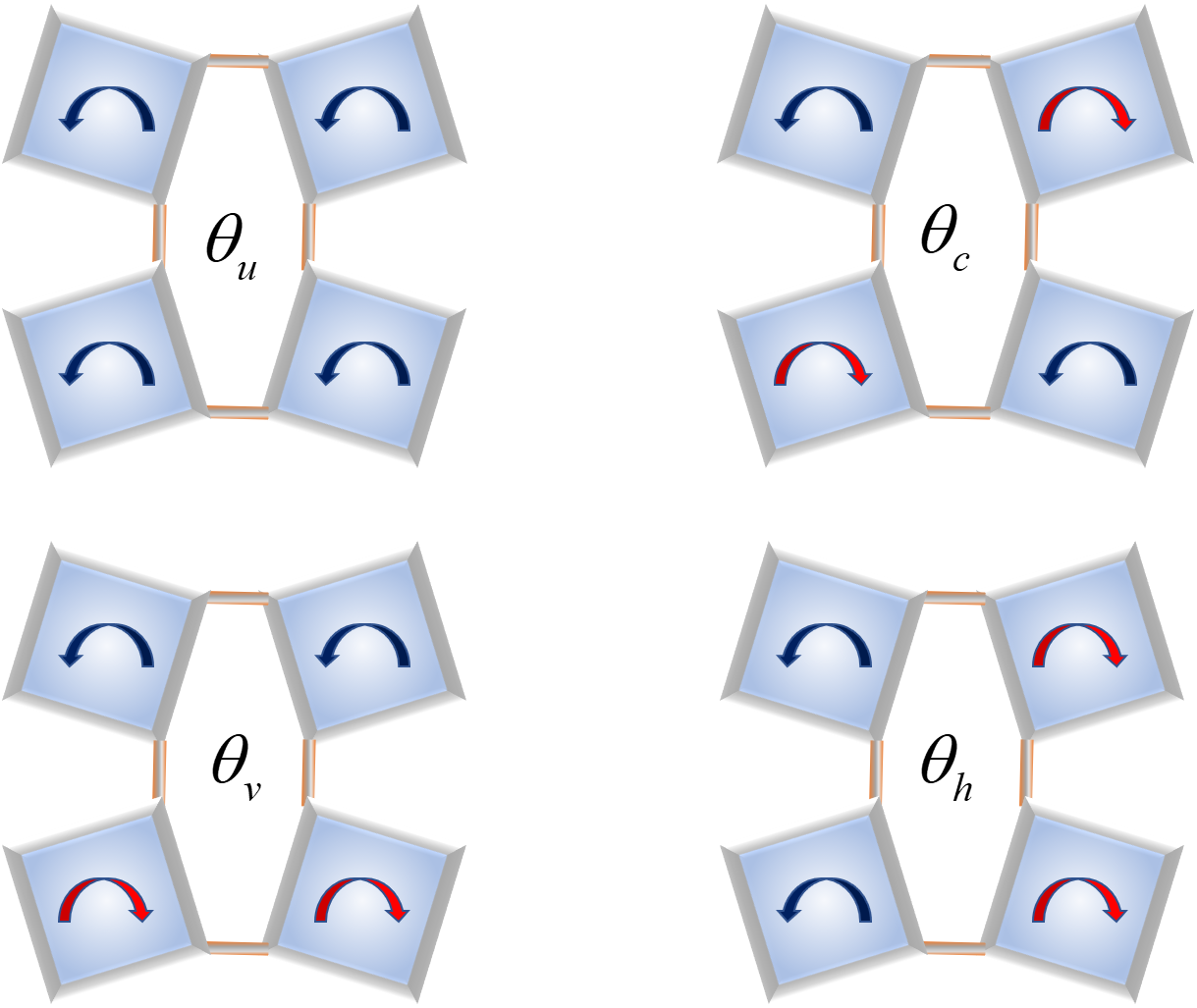}\protect\protect\caption{Collective rotational modes: uniform, checkerboard, vertical and horizontal.\label{fig:modes}}
\protect 
\end{figure}

 In absence of the additional symmetries in Grima's model, the system requires all the orientational degrees of freedom for its description; however, the previous scenario suggests that such description can be facilitated if written in terms of a judicious transformation of the angles of the subunits. Therefore, we introduce a new set of orientational modes:
\begin{equation}
\begin{pmatrix}\theta_u\\ \theta_c\\ \theta_v\\ \theta_h\end{pmatrix}=\frac{1}{2}
\begin{pmatrix}
1&1&1&1\\
1&-1&-1&1\\
1&1&-1&-1\\
1&-1&1&-1
\end{pmatrix}
\begin{pmatrix}\theta_{11}\\ \theta_{12}\\ \theta_{21}\\ \theta_{22}\end{pmatrix},
\end{equation}
where the subscripts $u$, $c$, $v$ and $h$ stand for \emph{uniform}, \emph{checkerboard}, \emph{vertical} and \emph{horizontal}, respectively (Fig. \ref{fig:modes}). The uniform mode, $\theta_u=\langle\theta\rangle_0$, can be thought of as a center of mass since it describes rotations of the structure as a whole. The checkerboard mode, $\theta_c=\langle\theta\rangle_1$, describes the coherent motion in which we are interested. And the horizontal and vertical modes complete the transformation. It can be seen that a uniform and checkerboard mode can be defined for any array with $2N\times 2N$ subunits.

To evaluate how the symmetry breaking introduced by the linkages affects the stability of coherent dynamics, we check the prevalence of the checkerboard mode in samples of 100 trajectories of 3000 time-steps with random initial orientations and random angular velocities, such that $\sum_{ij}\omega_{ij}^2\Delta t=0.04$, for arrays with several linkage lengths and various statistical dispersions for the random variables, but keeping $\kappa=0$.

\begin{figure*}[ht]
\protect\centering{}\protect\protect\includegraphics[scale=0.5]{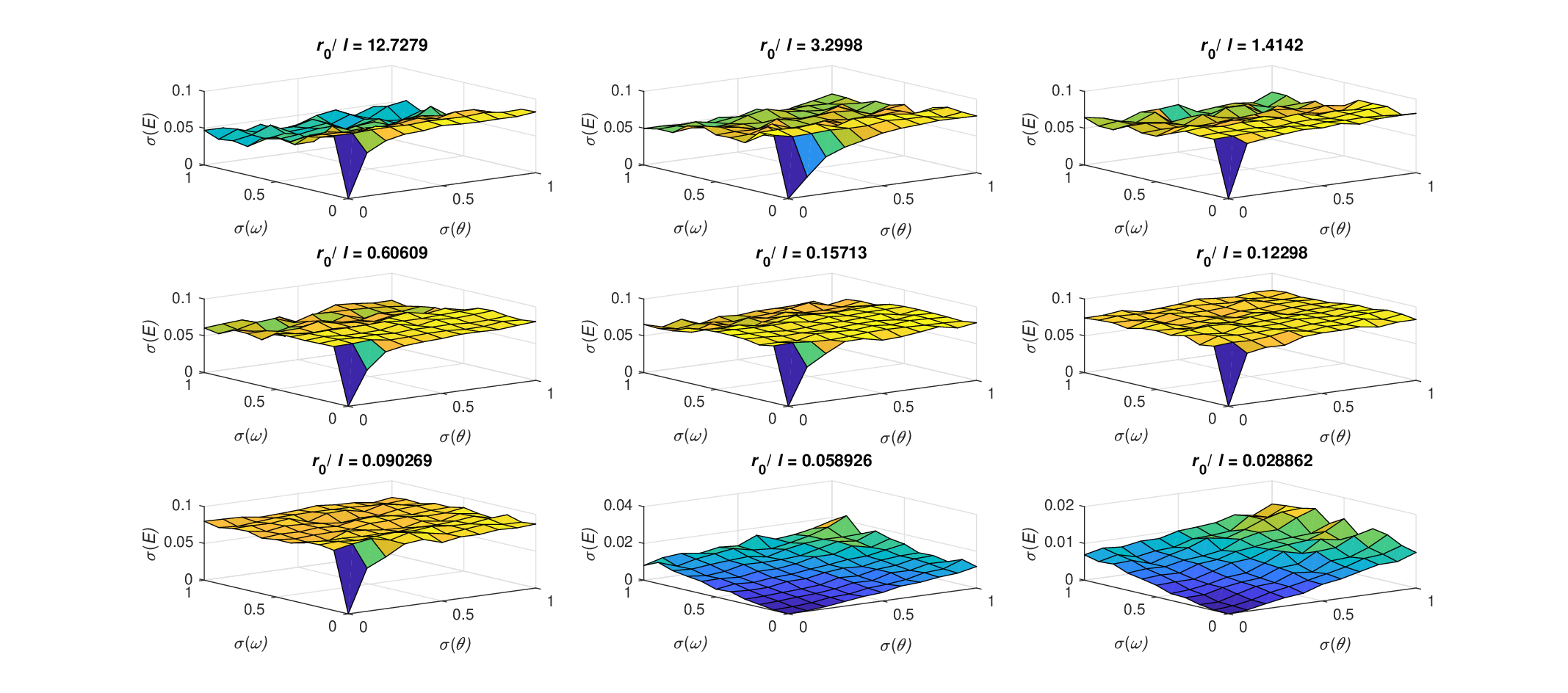}\protect\protect\caption{Standard deviation of the fraction of the total energy in the checkerboard mode as a function of the standard deviations in initial angle and initial angular velocity for various linkage lengths.\label{fig:components22}}
\protect 
\end{figure*}

Stability of coherent behavior would entail that the fraction of the total energy stored in the coherent mode remains constant during a trajectory. Under that rationale, Figure \ref{fig:components22} shows the standard deviation,$\sigma(E)$, throughout trajectories of the fraction of the total energy in the checkerboard mode, after removal of the uniform mode, as a function of the linkage lengths and standard deviations of the initial conditions, $\sigma(\theta)$ and $\sigma(\omega)$. The observed trends indicate that the contribution of the checkerboard mode becomes less relevant as the initial conditions are less reminiscent of it, and it never overtakes the dynamics again. The relevance of the mode also decreases as the linkages grow; this finding implies that the inclusion of the linkages reduces the likelihood of observing coherent dynamics.

\begin{figure}[ht]
\protect\centering{}\protect\protect\includegraphics[scale=0.23]{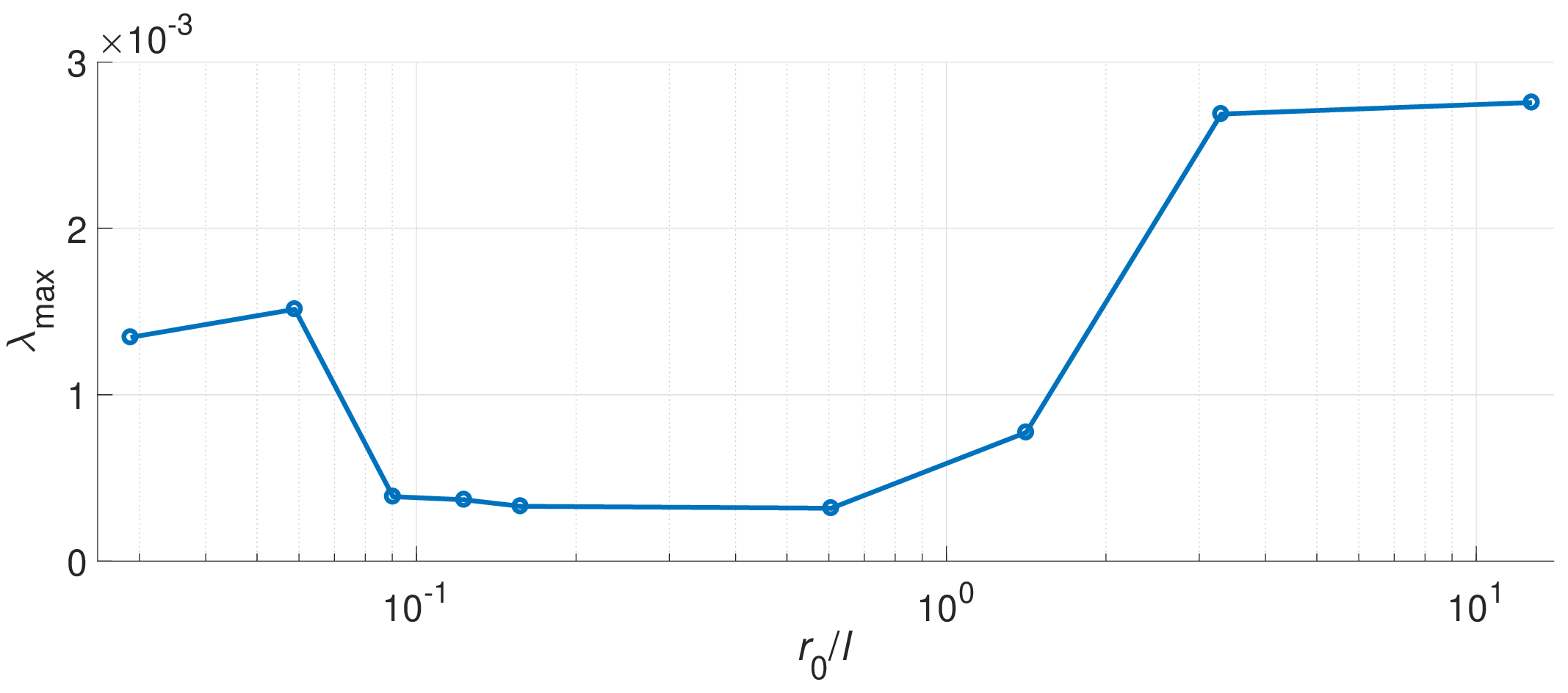}\protect\protect\caption{Maximal Lyapunov Exponent estimated for arrays of various linkage lengths.\label{fig:lyaps}}
\protect 
\end{figure}

\begin{figure}
\includegraphics[scale=0.25]{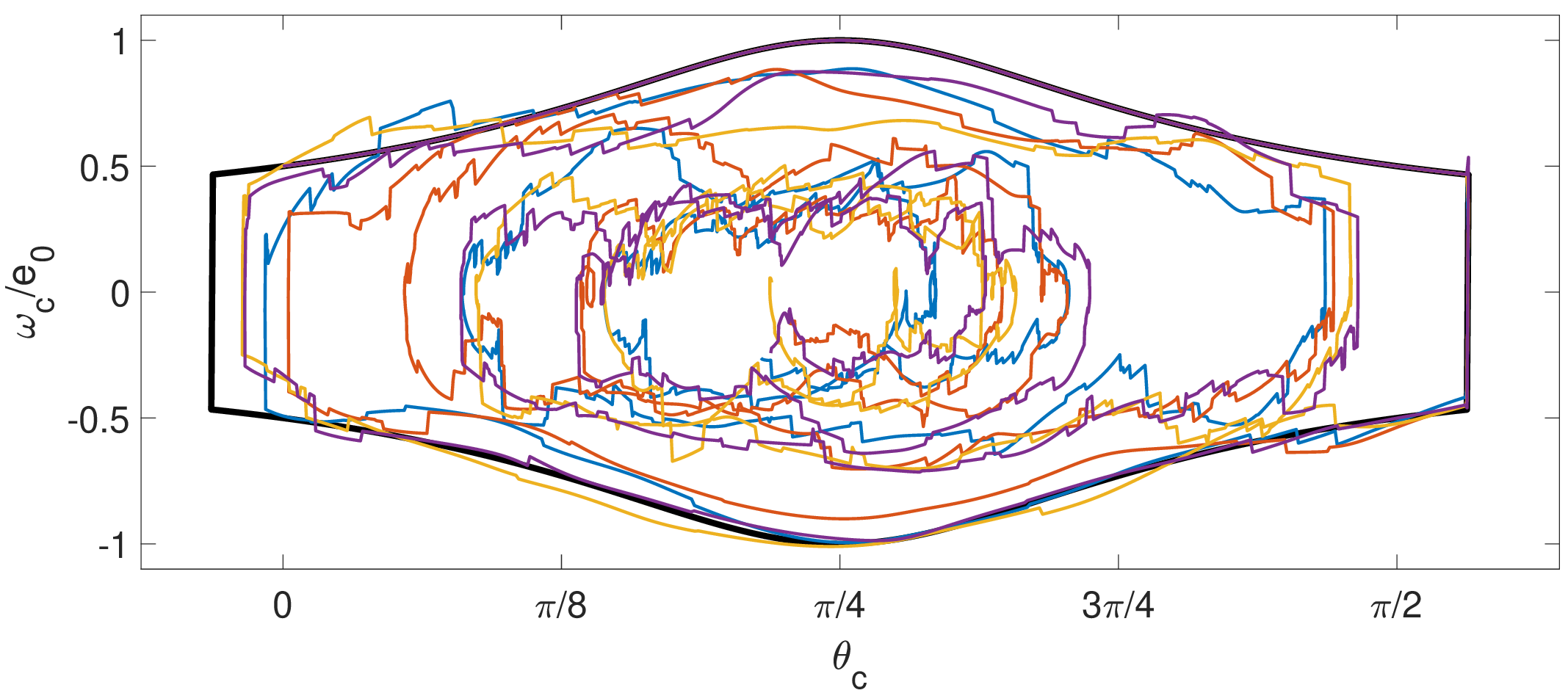}
\caption{Phase space with orbits initiated over a range of angular velocities. Divergence from the coherent orbit is evident.\label{fig:chaos}}
\end{figure}

To confirm the previous conjecture, we estimate the maximal Lyapunov exponent, $\lambda_{\max}$, of the system in the vicinity of the coherent orbit\cite{Wimberger2014} for various linkage lengths, maintaining $\kappa=0$, adapting the method in Refs. \citenum{Kantz1994} and \citenum{RosensteinCollinsDeLuca1993}. The results in Fig. \ref{fig:lyaps} reveal that the coherent orbit is unstable and the system presents a bias towards a chaotic behavior (Fig. \ref{fig:chaos}). As it could be expected, longer linkages produce dynamics with larger Lyapunov exponents; however, the dynamics of arrays with intermediate length linkages exhibit  smallest $\lambda_{\max}$.  This observation can be understood as a consequence of the increasing frequency of collisions for structures with shorter linkages. Nevertheless, the fact that the Lyapunov exponent remains positive regardless of the size proportions in the array strongly suggests that the coherence is not a consequence of geometry alone, and other interactions must play a role in preserving the single-domain morphology observed in RhuA samples.

\section{The 10$\times$10 structure\label{sec:10x10}}

The small structure discussed in the previous section lacks the constraining effects  whereby a high density of subunits in larger structures results in a higher frequency of collisional events.\cite{ChenKlotsaEngelEtAl2014} Thus, the question of whether the geometry of the structure on its own imposes coherent dynamics remains unanswered for larger arrays. We address this situations in the current section with a network of $10\times10$ subunits.

\begin{figure*}[ht]
\protect\centering\protect\protect\includegraphics[scale=0.5]{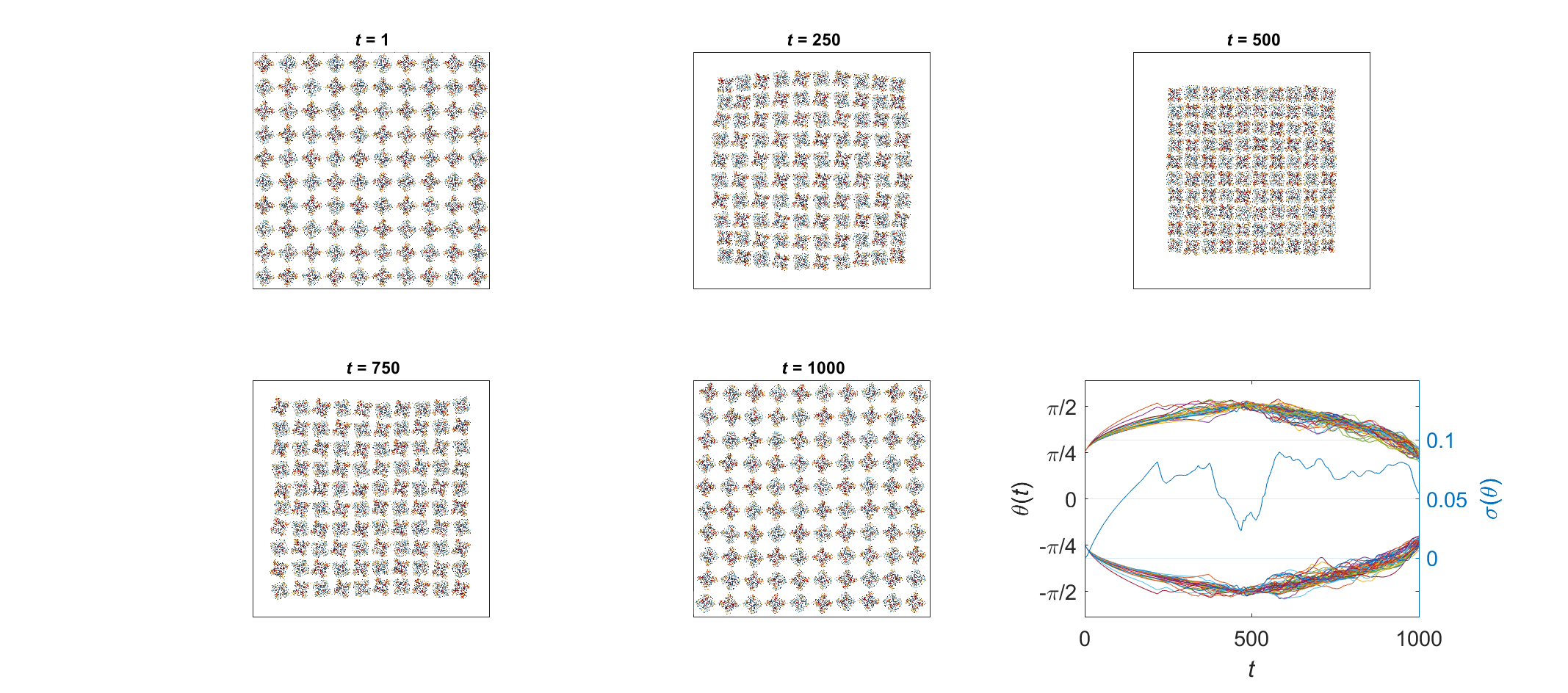}\protect\protect\caption{Panels 1-5: Snapshots at various time-steps of the simulated trajectory
for a $10\times10$ lattice with $r_{0}/l=5\%$, starting from the
open configuration and with uniform sign-altered initial angular velocities.\cite{GrueningerSchulz2008}\label{fig:snaps}
Bottom-right panel: behavior of the subunits orientation as a function
of time. The non-zero standard deviation indicates departure from coherent behavior.}
\protect 
\end{figure*}

Panels 1-5 of Fig. \ref{fig:snaps} show snapshots of the trajectory calculated starting from the open configuration with initial angular velocity only on the checkerboard mode such that $\omega_c(0)\Delta t=0.01$.

A remarkable observation from this example is the loss of the initial uniformity of the lattice during the advancement of the dynamics, which is illustrated in the last panel of Fig. \ref{fig:snaps}, where the angles of the subunits, as well as the standard deviation of their absolute value, are plotted as a function of time. The erratic behavior observed midway through the closed configuration can be explained by the fact that the number of positional constraints, \emph{i.e.}, the number of neighbors to which a subunit is connected, varies from two in the case of the corners, to three for squares at the edges, and four for those in the bulk. This observation illustrates how the introduction of non-zero length linkages undermines the capability of the system for absolute conformational coherence when compared to Grima's model,\cite{GrimaEvans2000} thus prompting further analysis.

To investigate the effect of the bond lengths on the coherence, we simulate several trajectories for a set of linkage lengths. For every given length, (pseudo)randomness of the trajectory sample space is achieved by running a trajectory starting from the open configuration with random initial angular velocities for the subunits, and randomly selecting 24 instants as seeds for new trajectories initialized with their own set of angular velocities for the subunits. The 24 trajectories thus obtained comprise the sample space per linkage length. The simulations, performed with resources from XSEDE,\cite{TownsCockerillDahanEtAl2014} were kept under 1000 time-steps due to two factors: for networks with short bonds, the statistical behavior does not feature, as we will show,  significant changes over that time window; while for the longest bonds, long trajectories reached nonphysical configurations, \emph{e. g.}, bonds intercrossing or cutting through subunits, which our current formalism does not properly avoid.

\begin{figure}[ht]
\protect\centering\protect\protect\includegraphics[scale=0.24]{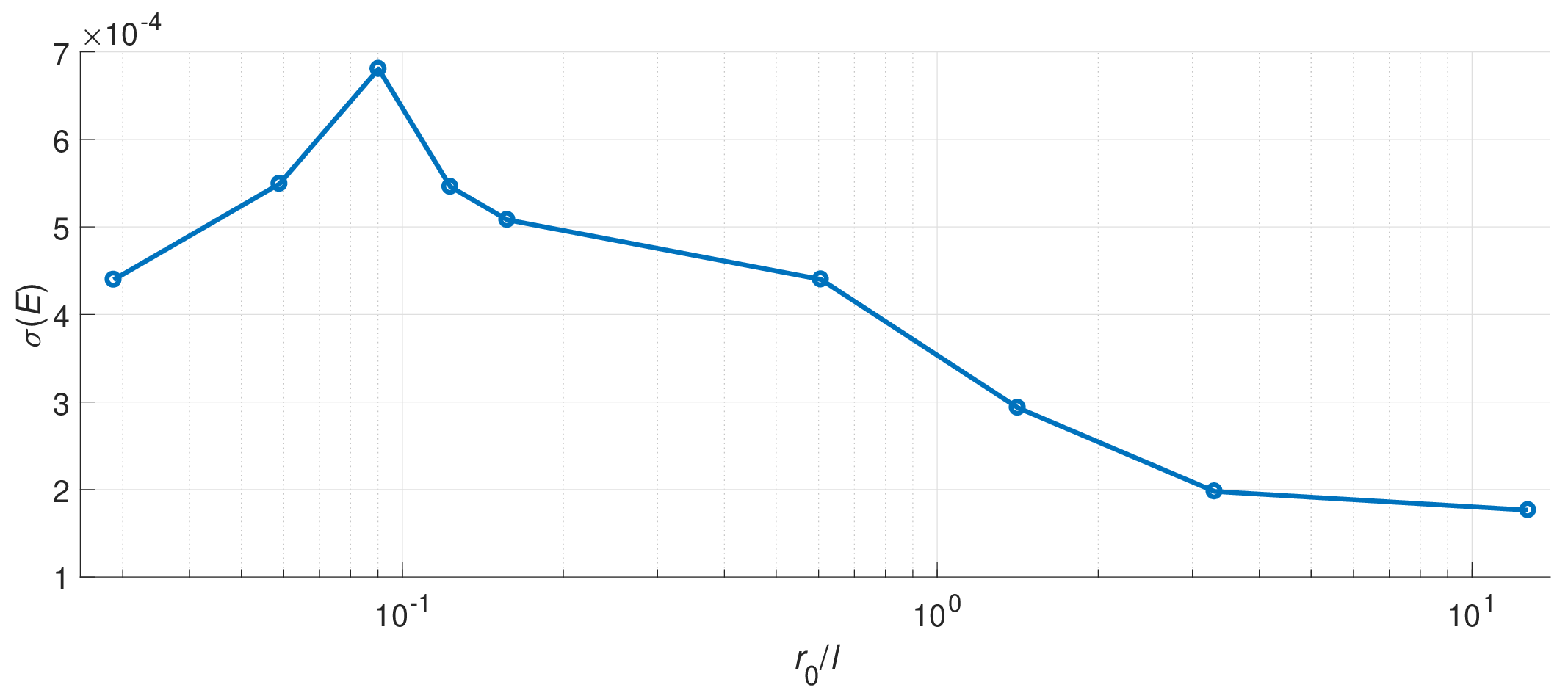}\protect\protect\caption{Trajectory standard deviations of the fraction of the rotational energy in the checkerboard mode for $10\times10$ arrays as a function of the linkage lengths.\label{fig:components10}}
\protect 
\end{figure}

Figure \ref{fig:components10} shows the statistical dispersion throughout trajectories, $\sigma(E)$, of the fraction of the total energy allocated in the checkerboard mode averaged over all the simulated trajectories after removing the uniform mode for arrays with several linkage lengths. We see that the dispersion is always non-zero, which implies that the checkerboard mode is never stable. Surprisingly, the expected behavior in which departure from Grima's model translates into less stability of the checkerboard mode is true only for short linkages since, after a certain point, the dispersion is observed to decrease. The fact that the checkerboard mode is more stable for longer linkages can be understood under the light that as the linkages grow, the frequency of collisions decreases, which means that the energy has less opportunity to spread among the other modes. This observation leads to the conclusion that collisions undermine the checkerboard mode, instead of stabilizing it.

\begin{figure}[ht]
\protect\centering\protect\protect\includegraphics[scale=0.24]{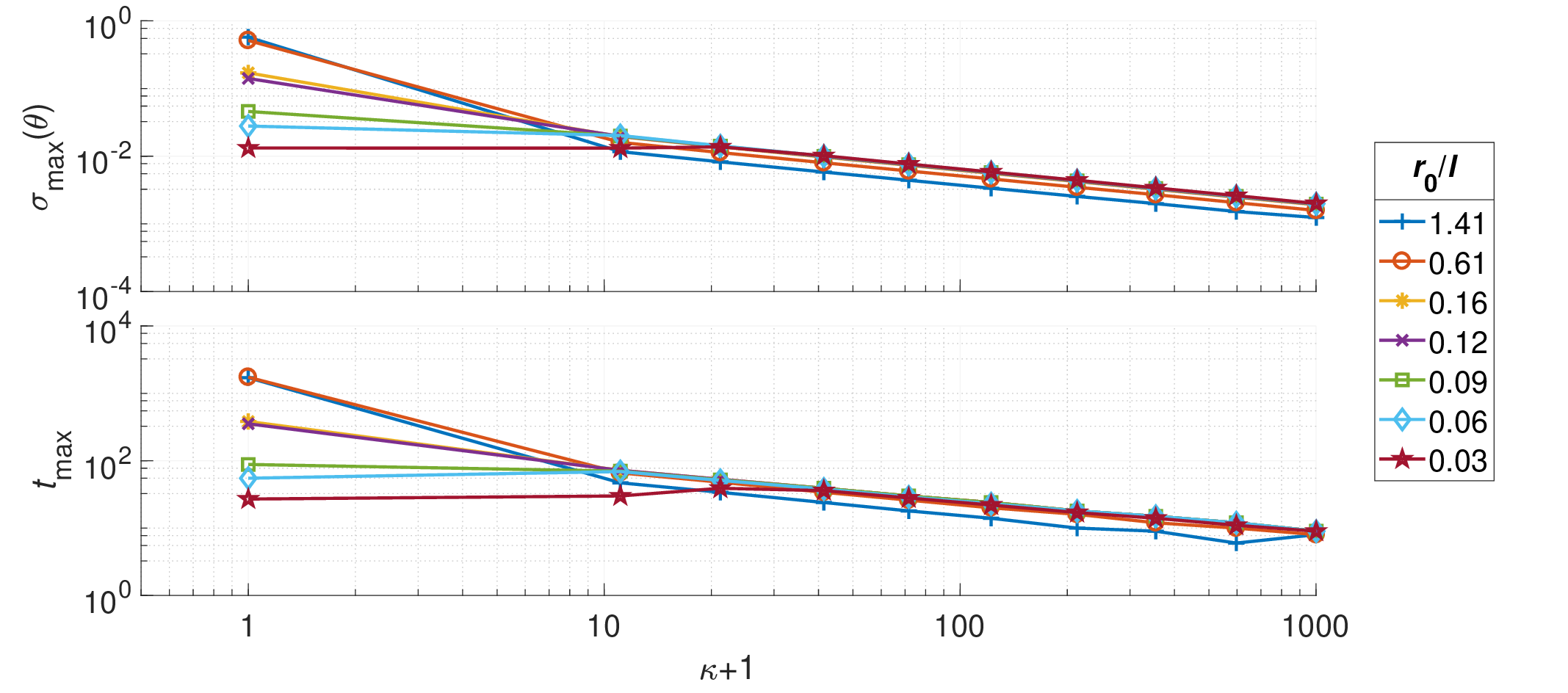}\protect\protect\caption{Maximum orientational standard deviation, and time to reach it, as functions of the coherence inducing constant.\label{fig:desvkap}}
\protect 
\end{figure}

Since there seems to be no evidence to support the notion that absolute coherence can emerge from the geometric features of the network alone, we proceed to evaluate the role of intermolecular forces that would correspond to interactions between the network and the solvent. In the current model, these interactions are represented by the potential $\sum_{ij}V_{ij}$ in Eq. \eqref{eq:lagrangian} and parametrized by the restitution constant $\kappa$. We investigate the effect of the potential by running, over a range of values of $\kappa$ and $r_0$,  trajectories initialized in the closed configuration and with angular velocity only in the checkerboard mode such that $\omega_c(0)\Delta t=0.01$, and registering the first maximum in the standard deviation of orientations, $\sigma_{\max}(\theta)$, as well as the time-step at which it occurs, $t_{\max}$. These results are summarized in Fig. \ref{fig:desvkap}.

When performing this analysis, the structures with linkages considerably bigger than the squares attained nonphysical configurations long before reaching the first maximum in orientational standard deviation; therefore, these arrays were excluded from the following discussion. For structures with linkages comparable in size to the squares and shorter, the dispersion and delay grow with the linkage length when $\kappa=0$, but non-zero force constants invert the trend. This observation agrees with the fact that bigger squares have a larger moment of inertia, thus making it harder for the coherence inducing force to change their velocities. However, the behavior of $\sigma_{\max}(\theta)$ for positive values of $\kappa$ seems to follow a negative power law that becomes less dependent on the proportions of the structure as the linkage gets shorter. As it can be seen, for trajectories with small values of $\kappa$, coherence is rarely retained since $\sigma_{\max}(\theta)$ is high and it requires many time-steps to be achieved. In contrast, as the force constant grows,  the maximum dispersion and its time of occurrence decrease.

The plots in Fig. \ref{fig:desvkap} reveal that absolute coherence is an asymptotic limit of the action of the force. Since the motion is inherently incoherent unless an extremely large potential forces it otherwise, we arrive at the same conclusion of Ref. \onlinecite{AlbersteinSuzukiPaesaniEtAl2018} that the presence of solvent molecules is essential in preserving the single-domain structure of these two-dimensional crystals along a conformational transition.

\section{Summary and Conclusions\label{sec:conclusion}}

We have developed an efficient and reliable procedure to simulate
the dynamical behavior of a network built from impenetrable squares
linked by rigid rods; this computational model emulates the geometry and topology of the auxetic two-dimensional crystals assembled from \textsuperscript{C98}RhuA, a system subject to constraints of both holonomic and nonholonomic nature. The constraints in velocity space are handled, regardless of their nature, by means of the MTM, which
we have shown to be a simple yet powerful tool. The treatment of constraints
in coordinate space is improved from the traditional approaches by applying the
exact values of the Lagrange multipliers (as taken from the acceleration
constraints) to initialize the iterative process. For shorter time-steps,
however, a constant-length ansatz provides a boost in performance
by eliminating the need for an iterative protocol. It is worth noting that this approach
is limited to short time-steps only because of the nonlinearity of
the angular variables considered in our model. For a system described
purely by Cartesian coordinates, the reliability of the method should
allow for longer step sizes. For instance, this method could be applied to solvent molecules for timescales in which they can be approximated as rigid structures.\cite{WychFraserMobleyEtAl2019}

Data extracted from the simulated trajectories produced using our developed
method were analyzed to characterize the dynamic behavior of the geometric
array. In particular, the uniformity of motion among building blocks was
evaluated as a function of the rigid-linkage lengths. We observe that absolute coherence is not achievable in general for finite linkage lengths, unlike in Grima's model, where the latter vanish. Furthermore, the only way for our model to detect behavior in which a single conformational domain is observed throughout the array is to include a delocalized effective potential with a non-negligible intensity. This is consistent with the findings in Ref. \citenum{AlbersteinSuzukiPaesaniEtAl2018} that the free energy of the solvent is appreciably sensitive to the configuration of the lattice.

The presented methodology establishes a framework to explore more complex structures, like hybrid or defective arrays, as well as a broad range of physical effects:  pairwise potentials and dissipative mechanisms, to name a few.

 \section*{Data Availability Statement}
The data that support the findings of this study are available from the corresponding author upon reasonable request.

\begin{acknowledgments}
The authors thank Robert Alberstein and Akif Tezcan for the details they provided about the experimental system and the network model that allowed the calibration of our timescale. The authors also thank Cecilia Clementi, Michael K. Gilson, Melvin Leok, Miguel A. Bastarrachea-Magnani, Octavio Narvaez-Aroche and Gerardo Soriano for their insightful comments and discussions. This work used the Extreme Science
and Engineering Discovery Environment (XSEDE) Oasis at the Comet Rapid
Access Project through allocation TG-ASC150024, which is supported
by National Science Foundation grant number ACI-1548562.
JCGA, RFR, and JYZ acknowledge support from the NSF Career Award CHE-164732.
JCGA also received funding from UC-MEXUS/CONACYT through scholarship ref. 235273/472318. GW was supported by the 2019 Undergraduate Summer Research Award from UC San Diego.
\end{acknowledgments}

\begin{appendix}

\section{Constant angular acceleration in terms of non-angular quantities\label{sec:angal}}
The kinematic EOM of a body in circular motion with radius $r$ are
\begin{align}
\vec{r}(t)=&r\left\{\cos[\phi(t)]\hat{e}_x+\sin[\phi(t)]\hat{e}_y\right\},\\
\dot{\vec{r}}(t)=&r\dot\phi(t)\left\{-\sin[\phi(t)]\hat{e}_x+\cos[\phi(t)]\hat{e}_y\right\},\\
\ddot{\vec{r}}(t)=&\frac{\ddot\phi(t)}{\dot\phi(t)}\vec{v}(t)-\left[\dot\phi(t)\right]^2\vec{r}(t).
\end{align}
The scalar products of acceleration with position and with velocity yield a system of equations of the form
\begin{equation}\label{eq:aplin}
\begin{cases}
\vec{r}(t)^T\ddot{\vec{r}}(t)=-[\dot\phi(t)]^2 r^2\\
\dot{\vec{r}}(t)^T\ddot{\vec{r}}(t)=\ddot\phi(t)\left\lvert\dot{\vec{r}}(t)\right\rvert^2/\dot\phi(t)
\end{cases},
\end{equation}
which can be solved for $\dot\phi(t)$ and $\ddot\phi$ to give
\begin{align}
\dot\phi(t)=&\frac{\left\lvert\vec{r}(t)^T \ddot{\vec{r}}(t)\right\rvert^{1/2}}{r},\\
\ddot\phi(t)=&\frac{\left\lvert\vec{r}(t)^T\ddot{\vec{r}}(t)\right\rvert^{1/2}\dot{\vec{r}}(t)^T\ddot{\vec{r}}(t)}{\left\lvert\dot{\vec{r}}(t)\right\rvert^2 r}
\end{align}
\end{appendix}

\bibliography{bibrhua}

\end{document}